%% file: paper1.tex
\documentclass[preprint2]{aastex60}
\input{setup}
\begin{document}
\title{Measuring Star-Formation Histories, Distances, and Metallicities with Pixel Color-Magnitude Diagrams I: Model Definition and Mock Tests}




\author{B.~A.~Cook\altaffilmark{1}, Charlie Conroy\altaffilmark{1}, Pieter van Dokkum\altaffilmark{2}, and Joshua~S.~Speagle\altaffilmark{1}}
\altaffiltext{1}{Center for Astrophysics | Harvard \&{} Smithsonian, 60 Garden St., Cambridge, MA 02138, USA}
\altaffiltext{2}{Astronomy Department, Yale University, 52 Hillhouse Ave, New Haven, CT 06511, USA}
\email{bcook@cfa.harvard.edu}

\begin{abstract}
We present a comprehensive study of the applications of the pixel color-magnitude diagram (pCMD) technique for measuring star formation histories (SFHs) and other stellar population parameters of galaxies, and demonstrate that the technique can also constrain distances.
SFHs have previously been measured through either the modeling of resolved-star CMDs or of integrated-light SEDs, yet neither approach can easily be applied to galaxies in the "semi-resolved regime".
The pCMD technique has previously been shown to have the potential to measure stellar populations and star formation histories in semi-resolved galaxies.
Here we present Pixel Color-Magnitude Diagrams with Python (\pcmdpy{}), a GPU-accelerated package that makes significant computational improvements to the original code and including more realistic physical models.
These advances include the simultaneous fitting of distance, modeling a Gaussian metallicity-distribution function, and an observationally-motivated dust model.
GPU-acceleration allows these more realistic models to be fit roughly $7\times$ faster than the simpler models in the original code.
We present results from a suite of mock tests, showing that with proper model assumptions, the code can simultaneously recover SFH, \FeH{}, distance, and dust extinction.
Our results suggest the code, applied to observations with \HST{}-like resolution, should constrain these properties with high precision within 10 Mpc and can be applied to systems out to as far as 100 Mpc.
pCMDs open a new window to studying the stellar populations of many galaxies that cannot be readily studied through other means.
\end{abstract}


\section{Introduction}
\label{s.intro}

The mass build-up of galaxies is sensitive to many physical processes.
Mergers, AGN activity, supernovae feedback, and the accretion of pristine gas from cosmological filaments will all affect a galaxy's star-formation history (SFH) and chemical enrichment, and it is a major goal of modern astrophysics to constrain their relative impacts.
Hydrodynamical simulations \citeeg{Hopkins2014, Vogelsberger2014a} are one powerful way to study the evolution of galaxies, but their results require reliable observational constraints on the diversity of SFHs and stellar populations observed in the universe.

To date, measurements of stellar populations and SFHs largely rely on two distinct techniques, each with limitations on the systems they can be applied to and the robustness of the results.
These approaches can be considered in a single framework through considering the typical number of stars per resolution element, \Npix{} \citep{vandokkum2014b, Conroy2016}.

The first technique, \textit{resolved star photometry}, typically requires $\Npix \lsim \mathcal{O}(10^{-1})$ in order to fully resolve each individual star in a system.
The fluxes of these stars in at least 2 bands are converted into a color-magnitude diagram (CMD), and the stellar populations are derived from fits to stellar evolution models \citeeg{Dolphin2002,Weisz2011,Lewis2015,Williams2015}.
As long as stars down to the oldest main-sequence turnoff are resolved, this method recovers highly precise and robust measurements of the SFH, and is considered the gold standard.

In practice, the number of systems where the oldest main-sequence turnoff can be resolved is small, even with the resolution of \HST{}'s Advanced Camera for Surveys (ACS). 
Several dozen dwarf galaxies in the local group have sufficiently low stellar densities to allow well-measured SFHs \citeeg{Weisz2011,Weisz2014}.
But even our nearest massive neighbor, M31, is so crowded that only rare, massive main-sequence stars \citep{Lewis2015} can be individually resolved in most fields, limiting SFH recovery to relatively recent star formation. In the inner regions of M31, with $\Npix \sim \mathcal{O}(10^1)$, SFH can be measured using the resolved red giant branch or red clump stars, but the results are subject to significant systematics due to disagreement between models of these phases of stellar evolution \citep{Williams2017}, with the oldest ages the most uncertain.
The inner bulge of M31 is so crowded that not even RGB stars can be resolved with \HST{}, making any SFH measurement impossible with this technique.

The second well-established method, \textit{SED modeling}, analyzes the integrated light of all stars in a (typically entirely unresolved) galaxy.
The broadband photometry and/or spectra are modeled using stellar population synthesis techniques to recover stellar populations and SFHs \citeeg{Walcher2011,Conroy2013b}. 
These methods typically assume a fully populated initial mass function (IMF), ignoring Poisson fluctuations of rare, evolved stars.
Such a regime requires $\Npix \gsim \mathcal{O}(10^6)$, usually a valid approximation for distant, unresolved galaxies.

Yet even if such fluctuations can be ignored, the light from the oldest, lowest-mass main-sequence stars is usually dwarfed by the light of rare, evolved stars, an effect known as "outshining", which can lead to an underestimate of the total stellar mass and oldest ages of star formation \citep{Maraston2010,Pforr2012a,Sorba2015}.
The SFHs recovered from SED modeling are also highly dependent on the underlying stellar evolution models \citeeg{Conroy2013b, Hunt2018}, with systematic uncertainties much larger than for resolved star SFHs.
\citet{Carnall2018b} and \citet{Leja2018b} also study the prior bias introduced from assuming various functional forms for the SFH.

In between the realm of resolved stars and integrated light, there is a substantial volume of the universe in the so-called \textit{semi-resolved regime} \citep{vandokkum2014b, Conroy2016}.
This regime can be defined loosely as $\mathcal{O}(10^1) \lsim \Npix \lsim \mathcal{O}(10^6)$, where stars cannot be individually resolved due to crowding, but surface-brightness fluctuations due to Poisson sampling of rare, bright stars in each pixel cannot be ignored.
This semi-resolved regime extends from $\sim$1 Mpc out to nearly 100 Mpc for massive galaxies observed with \HST{}.

The surface brightness fluctuations \citep[SBF; ][]{Tonry1988} distance technique is one example of a method for studying galaxies in the semi-resolved regime.
In this approach, distances to a galaxy are estimated by measuring the scale of the pixel-to-pixel surface-brightness fluctuations, studied in Fourier space in order to isolate the instrumental PSF.
The underlying stellar populations are largely treated as nuisance parameters and accounted for through calibration.
Other examples include the "disintegrated" light analysis of stellar halos \citep{Mould2012}, fluctuation spectroscopy \citep{vandokkum2014b}, and pixel-level time variability \citep{Conroy2015}.

A third technique for measuring SFHs and stellar populations, specifically applicable to semi-resolved galaxies, was introduced in \citet{Conroy2016}. The \textit{pixel color-magnitude diagram} (pCMD) method measures the fluxes in multiple (typically two) photometric bands within every pixel in an image, plotting the resulting magnitudes in a color-magnitude diagram.
The surface-brightness fluctuations across the image represent real variations in the number of rare, bright stars at each pixel.
Therefore, the distribution of these pixel fluxes (the pCMD) holds important information about the underlying stellar populations.
Critically, as we show in this work, pCMD distributions show distinct sensitivities to many key physical parameters, including metallicity, SFH, distance, and dust content. 
With a proper accounting of important observational artifacts, including PSF convolution, sky subtraction, and shot noise, these key parameters can be inferred through a forward-modeling procedure.

The pCMD technique can therefore be considered a generalization of SBF, as in this work we demonstrate that both distances and stellar populations can be simultaneously measured using the fluctuations in multiple bands of well-aligned photometry.
SBF is still an important rung on the distance ladder to this day, used to derive distances to nearby dwarf galaxies \citep{Cohen2018,Carlsten2019a} and massive galaxies alike \citep{Greene2019}. SBF also remains the best distance estimator to the host galaxy of GW170817, the first binary neutron star merger detected in gravitational waves \citep{Cantiello2018}.
Modeling the pCMDs of systems such as these offers the prospect of constraining (rather than assuming) their stellar populations, while more robustly accounting for their uncertainties in deriving distances.

This work presents the first results from Pixel Color-Magnitude Diagrams with Python (\pcmdpy{}), an open-source package developed to fully implement the pCMD fitting method and allow for modeling pCMDs with more realistic, flexible physical models.
The package extends significantly on the original framework outlined in \citet{Conroy2016}, allowing for a variety of metallicity, dust extinction, and star-formation history models, and for the first time extends the pCMD method to simultaneously fit for galactic distances.
The package is written in Python and hosted on GitHub\footnote{\url{https://github.com/bacook17/pcmdpy}, final API and documentation still in the development stages}, and posterior estimation is implemented with the dynamic nested sampling code \dynesty{} (Speagle in prep)\footnote{\url{https://github.com/joshspeagle/dynesty}}.
The computationally challenging simulation procedure has been accelerated using graphics processing units (GPUs), allowing for more accurate and complex models to be generated more rapidly and for fits to converge in less time.

This paper is outlined as follows. In \S{s.pcmdpy} we provide an outline of the \pcmdpy{} code.
In \S{s.tests} we describe a suite of mock tests where simulated pCMDs generated with the code are then fit to study the code's constraining power, and conclude the paper in \S{s.conclusions}.

\section{Pixel Color-Magnitude Diagrams with Python (\pcmdpy{})}
\label{s.pcmdpy}

\begin{figure*}
\plotonebig{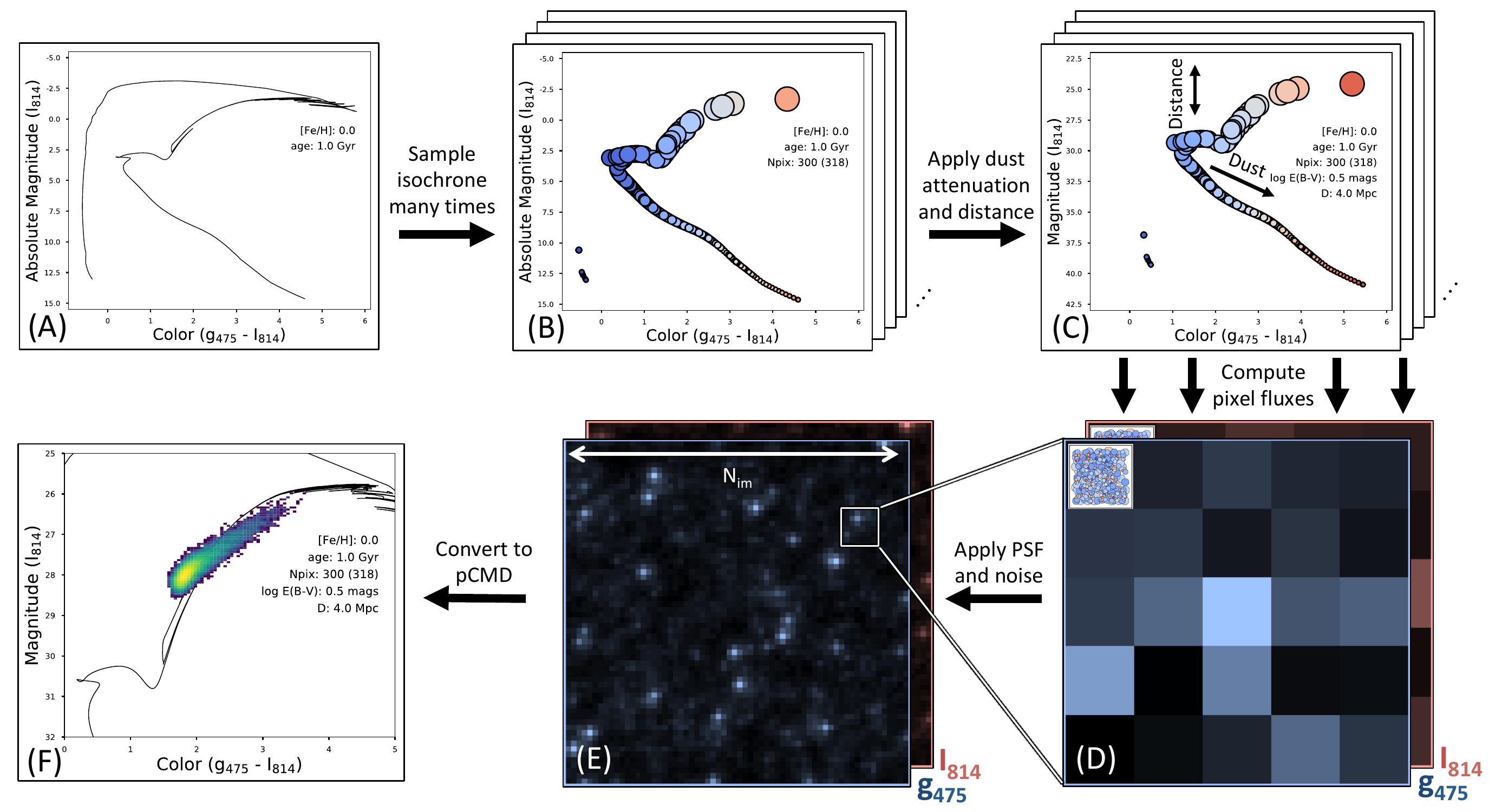}
\caption{Overview of the pCMD modeling procedure. (A) A metallicity and SFH model are chosen (for demonstration purposes, a single stellar population) and corresponding isochrones are drawn from MIST \citep{Choi2016}. (B) Stars are randomly sampled from the isochrones for each pixel in the simulated image, with mean number of stars in a pixel equal to \Npix{}. (C) The fluxes of each star are reddened due to dust extinction and adjusted by the distance modulus of the model. (D) The simulated images are generated, with each pixel flux equal to the sum of fluxes from all stars residing in that pixel, as shown in the top-left pixel. Here, images in two filters are shown. (E) The simulated images are convolved with the PSF models, and sky and shot noise are added. (F) The pixel color magnitude diagram is computed by converting pixel fluxes to magnitudes. The original isochrone track is shown for reference.}
\label{f.flowchart}
\end{figure*}

We describe here the procedure for generating and modeling a pCMD.
\S{ss.overview} discusses the general approach and primary assumptions made, while \S{ss.models} details the physical models implemented in \pcmdpy{}, including metallicity, star-formation history, distance, and dust extinction. 
\S{ss.sensitivity} demonstrates each free parameter in these models has a unique effect on pCMD distributions.
In \S{ss.infrastructure} we describe the computational architecture of the code.
\S{ss.sampling} describes our likelihood model for comparing pCMDs and the nested sampling approach for fitting model posteriors over these free parameters.

\subsection{Overview of the pCMD Technique}
\label{ss.overview}

A pixel color-magnitude diagram \citep[pCMD,][]{Conroy2016} represents the distribution in pixel-by-pixel photometry of a galaxy or galactic region, projected into magnitude space.
A pCMD is generated by measuring the flux in each pixel of two bands of photometric images and converting those fluxes to apparent magnitudes, representing their distribution in color-magnitude space as a Hess diagram.
When doing so, we are necessarily discarding any spatial correlations between the pixels.
This contrasts with the SBF method, where the Fourier transform preserves spatial information and can therefore isolate the effect of the PSF.
The distribution of the pCMD is thus shaped both by the physical properties (stellar populations) of the stars residing in the image and by the dust extinction and the optical properties of the telescope, such as the PSF and photometric noise.

The pCMD technique endeavours to constrain the underlying stellar populations through a \textit{forward modeling} approach: given a model for stellar photometry and our knowledge of the telescope's optical properties, we can create a simulated pCMD derived from a specified set of populations and compare it to real data.
This procedure is summarized visually in \F{f.flowchart}.
Comparing simulated pCMDs over a range of stellar populations allows us to infer the most likely populations to have generated in the data, as long as our stellar and observational models are reasonable approximations of the truth.

In practice, the possible combinations of stars residing in a given image is limitless, requiring us to make simplifying assumptions for the problem to be tractable.
We adopt a probabilistic approach: we assume there is a global, underlying distribution of stars within the image, from which each pixel represents a unique random realization. 
Therefore, we care only about the overall distribution of stellar populations and pixel magnitudes (the pCMD), not about matching each specific pixel to a model of the stars it contains.

Specifically, we assume that the number of stars in a pixel is drawn from a Poisson distribution, with mean number of stars constant across the image and equal to the free-parameter $\Npix$. Therefore, the surface-brightness fluctuations across pixels are a true signal, representing the Poisson noise in stars per pixel, the magnitude of which is determined by \Npix{} and the stellar populations.
\Npix{} can be estimated from the column density $\mathrm{N}_\mathrm{col}$ of stars, the observed spatial resolution $\theta$, and the distance $\mathrm{d}$ to the source, and is approximately:
\begin{align}
\Npix &\approx 6 \frac{\mathrm{stars}}{\mathrm{pixel}} \left(\frac{\mathrm{N}_\mathrm{col}}{10^3 \mathrm{pc}^{-2}}\right) \left(\frac{\theta}{0.05\arcsec}\right)^2 \left(\frac{\mathrm{d}}{1 \mathrm{Mpc}}\right)^2\,.
\label{eq.Npix}
\end{align}
\changed{The representative values above assume a galaxy with average stellar density of 1 star per cubic parsec and a thickness of 1 kpc ($\mathrm{N}_\mathrm{col} = 10^3 \mathrm{pc}^{-2}$), and observations with the spatial resolution of \HST{}-ACS. }

To determine the properties of those $\approx\Npix$ stars in each pixel, we assume a model for the distribution of their ages (the star-formation history model, SFH), metallicities (given as iron abundance, \FeH), and initial masses (an initial mass function, IMF). This is shown in Panel A of \F{f.flowchart}.

In our approach, we divide the age-\FeH-mass space into discrete bins, assigning each point a weight according to the models above and normalizing the weights to equal \Npix.
For every pixel in the simulated image (of size $\Nim\times\Nim$ pixels), we randomly draw the number of stars in each bin according to a Poisson distribution with the given weights\footnote{Given many Poisson-distributed numbers $n_i$ with weight parameters $\lambda_i$, the sum $N=\sum_i n_i$ is also Poisson-distributed, with weight parameter $\lambda = \sum_i \lambda_i$.}.
We then derive the absolute magnitudes of each star in the observed filters from stellar evolution models (isochrones), although this could also be implemented with well-calibrated observational catalogues. This is shown in Panel B of \F{f.flowchart}.

The intrinsic fluxes of each star are attenuated by dust extinction, potentially by both foreground (Milky Way) and source (host galaxy) dust, requiring reddening curves and an assumed model for the dust abundance.
The stellar magnitudes are converted to fluxes according to the distance to the host galaxy and are summed into their respective pixels. These steps are Panels C and D of \F{f.flowchart}.

Finally, the telescope and instrumental signatures must be accounted for (Panel E of \F{f.flowchart}).
The raw images are convolved with models for the point-spread function (PSF) in each filter\footnote{The PSF convolution breaks the assumption that each pixel represents an independent draw from an underlying distribution, as the fluxes in neighboring pixels are highly correlated. This makes writing down an exact likelihood or computing it using probabilistic programming intractable.}.
Sky flux can be added at this stage.
Poisson shot noise is added, assuming the data are in electrons, such that noise $= \sqrt{\mathrm{counts}}$. The resulting pCMD is computed by converting the pixel counts to magnitudes (Panel F of \F{f.flowchart}).

\changed{We additionally note that it is possible to both measure and simulate pCMDs in any arbitrary number of photometric bands.
In this work we only consider the case of two filters, primarily due to the challenge of comparing pCMD distributions in more than 2 dimensions.}

\subsection{\pcmdpy{} Model Choices}
\label{ss.models}


The \pcmdpy{} physical model for a region contains four primary model components: a metallicity model, a SFH model, a dust extinction model, and a distance model.
When fitting a pCMD, the free parameters of the fit all correspond to one of these four components, and we assume flat priors over each.
Each component (with the exception of distance) could be modeled in a variety of complex ways, and we detail here the implementations available in \pcmdpy{}.
\changed{The functional forms included here may not necessarily be adequate representations of the true shapes of the distributions of interest in particular galaxies, but they are designed to provide reasonable approximations.
In \S{ss.mismatch}, we discuss the effects of incorrect model assumptions.}

All of these physical model components are summarized in \T{t.models}. All other parameters or constants assumed in the model are considered hyper-parameters, and are held fixed throughout the fit. These are summarized in \T{t.hyperparams}.

\begin{table*}
\begin{center}
\begin{tabular}{|c|l|c|c|}\hline
No. & Model Name & Free Parameters & Hyper-parameters \\\hline\hline
\multicolumn{4}{|c|}{Metallicity Models}\\\hline
M1 & Single \FeH{}  & \FeH{} & \\\hline
M2 & Gaussian MDF (Fixed-Width)  & \mufeh{} & \sigfeh{} \\\hline
M3 & Gaussian MDF  & \mufeh{}, \sigfeh{} & \\\hline
\multicolumn{4}{|c|}{Star Formation History Models}\\\hline
S1 & Single Stellar Population  &$\log$\Npix{}, $\log \mathrm{age}$ & \\\hline
S2 & Constant SFR  & $\log$\Npix{} & \\\hline
S3 & Tau Model  & $\log$\Npix{}, $\tau$ & \\\hline
S4 & Delayed-Tau Model  & $\log$\Npix{}, $\tau$ & \\\hline
S5 & Non-Parametric SFH & $\log\mathrm{SFH}_{0}, \log\mathrm{SFH}_{1}, \cdots$ & $\mathrm{N_{SFH}}$, Bin Edges \\\hline
\multicolumn{4}{|c|}{Dust Extinction Models}\\\hline
E1 & Fixed Dust Screen &  \logEBV{} & \Fdust{} \\\hline
E2 & Log-Normal Dust Screen (Fixed-Width) & \mudust{} & \Fdust{}, \sigdust{} \\\hline
E3 & Log-Normal Dust Screen & \mudust{}, \sigdust{} & \Fdust{} \\\hline
\multicolumn{4}{|c|}{Distance Models}\\\hline
D1 & Fixed Distance & & \dmod \\\hline
D2 & Variable Distance & \dmod & \\\hline
\end{tabular}
\caption{Modeling a pCMD requires an assumed physical model for each of: metallicity, star-formation history, dust extinction, and distance. The physical model implementations in \pcmdpy{} and their parameters are listed here. Free-parameters are fit to data, while hyper-parameters are set prior to fitting.}
\end{center}
\label{t.models}
\end{table*}

\begin{table*}
\begin{center}
\begin{tabular}{|l|c|p{8cm}|}\hline
Parameter Name (Default) & Num Params & Description\\\hline\hline
Filters & \Nbands{} & Specify which observational filters to simulate. Includes zero-points and PSF models. \\\hline
\Nim{} (512) & 1 & Size of simulated image-plane (\Nim{} pix $\times$ \Nim{} pix)\\\hline
\Nage (21) & 1 & Number of isochrones to draw from in a SFH\\\hline
IMF (Salpeter) & 1 & The stellar initial-mass function \\\hline
Sky Level (0) & \Nbands{} & Level of background sky noise to add to each band\\\hline
Exposure Time & \Nbands{} & Exposure time of each image (important for modeling shot noise)\\\hline
Downsampling (5) & 1 & Factor to downsample isochrone points\\\hline 
Hess Binning (0.05, 0.05) & 2 & Width of Hess bins used to compute Log-Likelihood \\\hline
\end{tabular}
\end{center}
\caption{Global hyper-parameters of the \pcmdpy{} modeling procedure. These parameters remain fixed throughout fitting. Does not include parameters of the sampling algorithm (see Dynesty documentation).}
\label{t.hyperparams}
\end{table*}

Three metallicity models are available, which specify the metallicity-distribution function (MDF, in terms of \FeH{}) of the stars in each pixel. 
\begin{enumerate}
    \item[M1.] \textbf{Single [Fe/H]}: All stars have the same metallicity, equal to the single free parameter \FeH{}.
    \item[M2.] \textbf{Gaussian MDF (Fixed-Width)}: Stellar metallicities are drawn from a Gaussian distribution, with one free parameter corresponding to the mean (\mufeh{}). The standard-deviation (\sigfeh{}) is a hyper-parameter and is held fixed.
    \item[M3.] \textbf{Gaussian MDF}: As in M2, but \sigfeh{} is a second free parameter.
\end{enumerate}

We adopt the non-rotating isochrones of the MIST project, v1.2 \citep{Choi2016}.
The metallicities provided in the MIST isochrones are discretized to a grid with spacing 0.25 dex.
For model M1, we interpolate the isochrones between these grid points to recover metallicities not on the grid.
For models M2 and M3, we use only the metallicities on the grid, and weight their abundances by the specified Gaussian distribution.
Grid points with weights less than $1\%$ are removed, resulting in 5 to 10 metallicity points for most Gaussian models.

Five star-formation history models are implemented, which specify the distribution of stars as a function of age.
The number of stars per pixel \Npix{} is either an explicit free-parameter of the model, or computed as the sum of the SFH.
\begin{enumerate}
    \item[S1.] \textbf{Single Stellar Population}: All stars are of the same age. The two free parameters are the age of stars (in log years) and \Npix.
    \item[S2.] \textbf{Constant SFR}: Assumes constant star-formation rate (SFR) at all times. The one free parameter is \Npix{}, to which the total SFH sums.
    \item[S3.] \textbf{Tau SFH}: An exponentially-falling SFR ($\mathrm{SFR} \propto e^{-t / \tau}$). The two free parameters are $\tau$ (in Gyr) and \Npix{}.
    \item[S4.] \textbf{Delayed-Tau SFH}: A linearly-rising SFR followed by an exponential falloff ($\mathrm{SFR} \propto t e^{-t / \tau}$). The two free parameters are $\tau$ (in Gyr) and \Npix{}.
    \item[S5.] \textbf{Non-Parametric SFH}: The star-formation history is binned into several (by default: 5) independent bins, within which the SFR is constant. The free parameters are $\{\log\mathrm{SFH}_i\}$, the logarithm of total star-formation in each bin, in units of stars-per-pixel. The number and edges of the bins are hyper-parameters.
\end{enumerate}
With the exception of model S1, stellar ages are discretized using a grid of ages.
By default, the grid uses 21 equally-spaced bins in log age from 1 Myr to 14 Gyr, but this is a hyper-parameter \Nage{} that can be adjusted.
The age points are taken as the midpoints of each bin ($10^{6.1}$, $10^{6.3}$, $\cdots$, $10^{10.1}$ years), and the star-formation rate is assumed to be constant within the bin.

When simulating complex models represented by many metallicity and age points (ex: M2+S3), we downsample the MIST isochrones in mass by a factor of $5$ (a hyper-parameter).
This downsampling factor improves computation time dramatically while not significantly affecting the resulting pCMDs, as confirmed through internal tests.

In the current implementation of \pcmdpy{}, SFH and metal abundance are modeled independently: the metallicity of a star does not depend on its age.
In reality, the abundances of stars are known to evolve with age as previous generations enrich the ISM out of which stars form.
Future work could model both jointly, but this is outside the scope of the current work. 

We adopt a Salpeter IMF \citep{Salpeter1955} by default, but a Kroupa IMF \citep{Kroupa2001} is also implemented in \pcmdpy{}.
In addition to determining the distribution of stars by mass, the IMF also determines the conversion from \Npix{} to $\mathrm{M_{pix}}$, the stellar mass formed in each pixel.
The total stellar mass in the image, $\mathrm{M_\star}$, requires a solution for SFH to account for stellar mass loss from post-MS stellar evolution. 

Three dust extinction models are implemented, which determine the dust extinction in units of the reddening parameter, $\logEBV{}$. 
We adopt the framework of \citet{Dalcanton2015}, who studied of the dust extinction in M31. See their Section 3 for details.

\begin{enumerate}
    \item[E1.] \textbf{Constant Dust Screen}: All pixels have a constant amount of extinction, equal to the one free parameter: \logEBV{}.
    \item[E2.] \textbf{Log-Normal Dust Screen (Fixed-Width)}: The dust extinction in each pixel is drawn from a log-normal distribution, with one free parameter: the median extinction \mudust{}. The dimensionless width-parameter \sigdust{} is a hyper-parameter and is held fixed.
    \item[E3.] \textbf{Log-Normal Dust Screen}: As in E2, except \sigdust{} is an additional free parameter.
\end{enumerate}

Each dust model assumes a single, thin screen of dust. The geometry of the screen is specified by a hyper-parameter \Fdust{}, which determines the fraction of stars that are reddened by dust.
$\Fdust=1.0$ represents a foreground screen of dust (all stars are reddened), while our default choice of $\Fdust=0.5$ corresponds to a mid-plane disk of dust, with half the stars reddened and half unobscured.
This model assumes that in practical applications, any foreground dust from the Milky Way can be accounted for in data reduction.
Future work could extend to more complex dust geometries, as preliminary tests indicate neither of these models may be sufficient to model the complex and dense dust lane structures found in disk galaxies at the scale of interest.
We convert $\logEBV{}$ to magnitudes of extinction in each optical band using the $R_V = 3.1$ reddening law from \citet[][their table 6]{Schlafly2011}.

The final physical model component is distance. We implement two distance models, simply representing whether or not distance (in units of distance modulus, $\dmod$) is included in the fit:
\begin{enumerate}
    \item[D1.] \textbf{Distance Fixed}: Distance is assumed known. $\dmod$ is a fixed hyper-parameter.
    \item[D2.] \textbf{Distance Free}: $\dmod$ is a free parameter.
\end{enumerate}

A physical model is specified with one of each of these model components.
We will occasionally refer to the entire model used to simulate a pCMD using the alpha-numerical codes above. For example, our "fiducial-$\tau$" model (\S{ss.mocks}) is described as M2+S3+E2+D2.
This translates to a Gaussian MDF (with fixed $\sigma_{\FeH}$), a Tau SFH, a Log-Normal dust screen (with fixed $\sigdust$), and distance free.

A major hyper-parameter of the fitting technique is \Nim{}, the 1D size of the simulated image plane (the total number of pixels in the simulated pCMD is therefore $\Nim^2$).
$\Nim$ does not have to match the size of the data being fit, as we can compare the relative number of pixels in the pCMD.
Larger simulated images are more computationally expensive but reduce the inherent stochasticity of the likelihoods (see \S{ss.sampling}) by providing more samples of the rare fluctuations in surface-brightness.
We choose a default image size of $\Nim=512$, which we find to be the optimal size for reducing stochasticity while allowing for convergent fits in a reasonable time. 

We must match the observational conditions under which the data were taken as closely as possible, requiring several well-calibrated hyper-parameters representing each observed filter.
This includes the exposure time in each filter, the photometric zero-points, and a model for the PSF.
We have specifically modeled observations with the \HST{}-\textit{ACS} camera, but the model can be generally applied to ground-based observations as well.
The exposure time is taken from the FITS header of the imaging data.
Zero-points are computed using the \texttt{PySynphot} \citep{Lim2015} package, as detailed in the ACS Handbook\footnote{\url{http://www.stsci.edu/hst/acs/analysis/zeropoints}}.

PSF models are taken from the Tiny Tim \citep{Krist2011} web interface\footnote{\url{http://www.stsci.edu/hst/observatory/focus/TinyTim}}.
To simulate sub-pixel PSF effects, we subdivide each image into a $4\times{}4$ grid and apply to each a different PSF convolution, shifted by fractions of a pixel.
Tests showed our approach is a reasonable approximation to the effects of truly sub-sampling each pixel and applying a sub-pixel PSF model, and is substantially faster.

\subsection{Sensitivity of pCMDs to Model Parameters}
\label{ss.sensitivity}

\begin{figure*}
\plotonebig{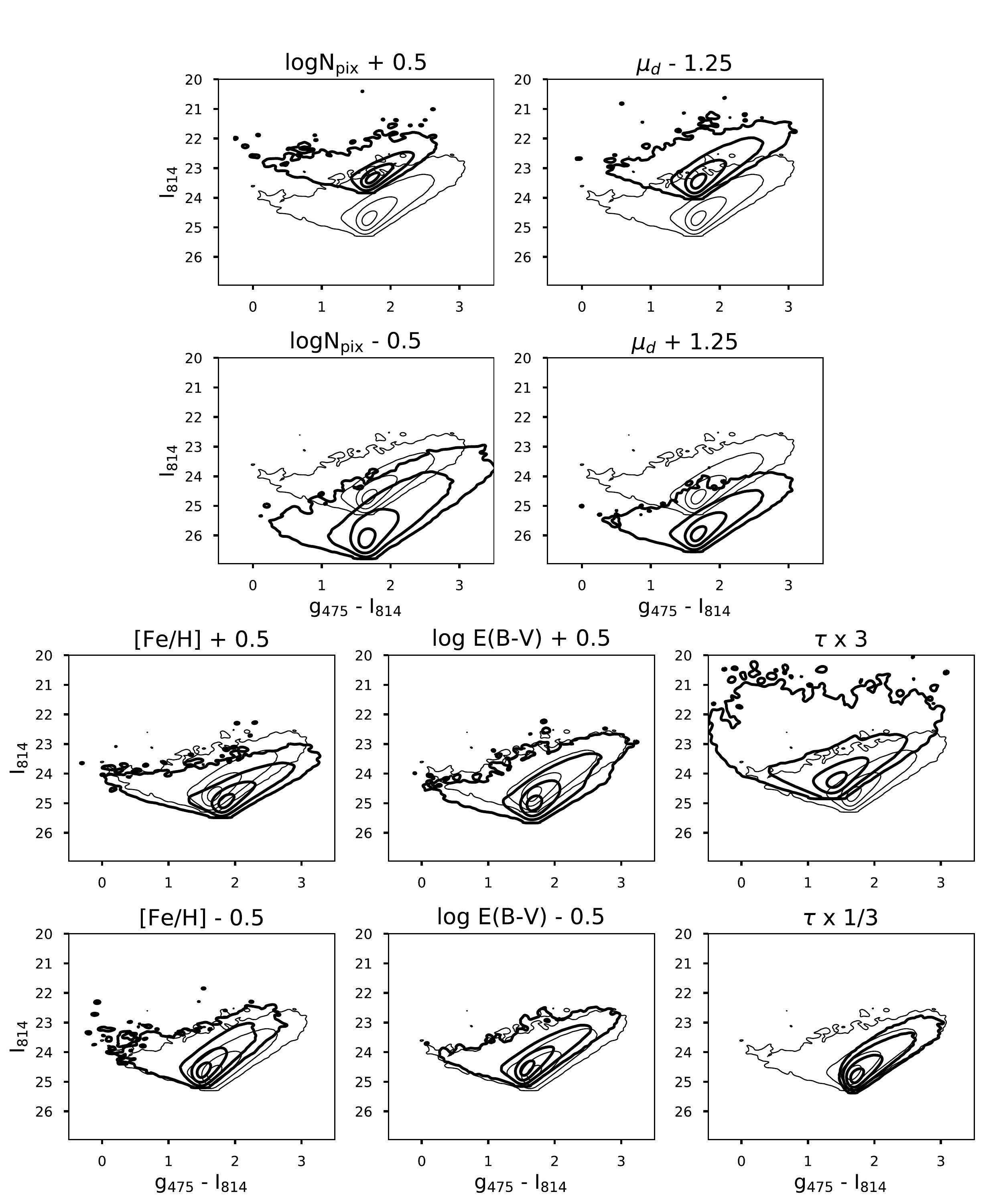}
\caption{Sensitivity of the pCMD distribution to the various model parameters. The baseline model, shown in thin grey, is a fiducial-$\tau$ model (see \S{ss.mocks}). The contours show the bounds within which 39\%, 87\%, 99\%, and 99.9\% of the points lie (the $1\sigma$, $2\sigma$, $3\sigma$, and $4\sigma$ contours, respectively). Varying each model parameter, shown in dark black, has \changed{significant} effects on the shape and location of the pCMD. In the top two rows, the changes in \Npix{} and \dmod{} were chosen such that the average flux is equal in each row.}
\label{f.params}
\end{figure*}

In \F{f.params}, we show that the detailed structure of pCMDs are sensitive to each of the physical parameters of interest, \changed{in} ways that allow constraints on each from only two bands of semi-resolved photometry.
In each panel, we show a simulated pCMD for a "baseline model" in grey, with a comparison for a model with one physical parameter changed superimposed in black. The baseline model has a single-\FeH{} metallicity model (M1, $\FeH = -0.5$), a Tau SFH (S3, $\Npix = 10^3,\;\tau=3$ Gyr), constant dust screen (D1, $\logEBV{} = -0.5$), and a distance of 1 Mpc ($\dmod = 25$). 

Of particular interest are the effects of \Npix{} (proxy for stellar surface density) and distance.
The upper-left column of \F{f.params} shows that in addition to changing the average luminosity, \Npix{} also affects the dispersion of the pCMD distribution, due to the increase in Poisson surface-brightness fluctuations.
We compare this to variations in distance modulus in the upper-right column, which simply shifts the distribution in the vertical (luminosity) direction, with no effect on color.
We show variations in \Npix{} and \dmod{} that each result in the same average luminosity; the dispersion of the distribution can be used to break the degeneracy between luminosity and distance.
This represents a reformulation of the SBF method: as the same physical system is moved towards larger distance, surface-brightness remains constant but the magnitude of fluctuations decreases as \Npix{} rises.
This hints at the utility of pCMDs to simultaneously recover distances and stellar populations for galaxies in the semi-resolved regime.
We demonstrate this conclusively in \S{s.tests}.

Varying metallicity shifts the peak of the distribution and has a notable effect on the slope of the upper-right wing.
The pixels in that region contain red-giant branch (RGB) stars, and the pCMD is therefore sensitive to the metallicity-dependent slope of the RGB \citep{Choi2016}.
The effects of increasing dust broadens the overall distribution but leaves the slope of the RGB-feature relatively intact.
As we show in \S{ss.results}, the model is able to constrain dust and \FeH{}, although there remains a slight degeneracy between the two.

Changing the age parameter, $\tau$, has significant effects on the blue-wing of the pCMD. This represents the relative abundance of young, massive main-sequence stars.
These hot stars are only abundant when there has been recent star-formation (high $\tau$), while only the old main-sequence and RGB stars contribute when recent star-formation is suppressed (low $\tau$).

\subsection{Computational Infrastructure}
\label{ss.infrastructure}

\begin{figure}
    \plotonebig{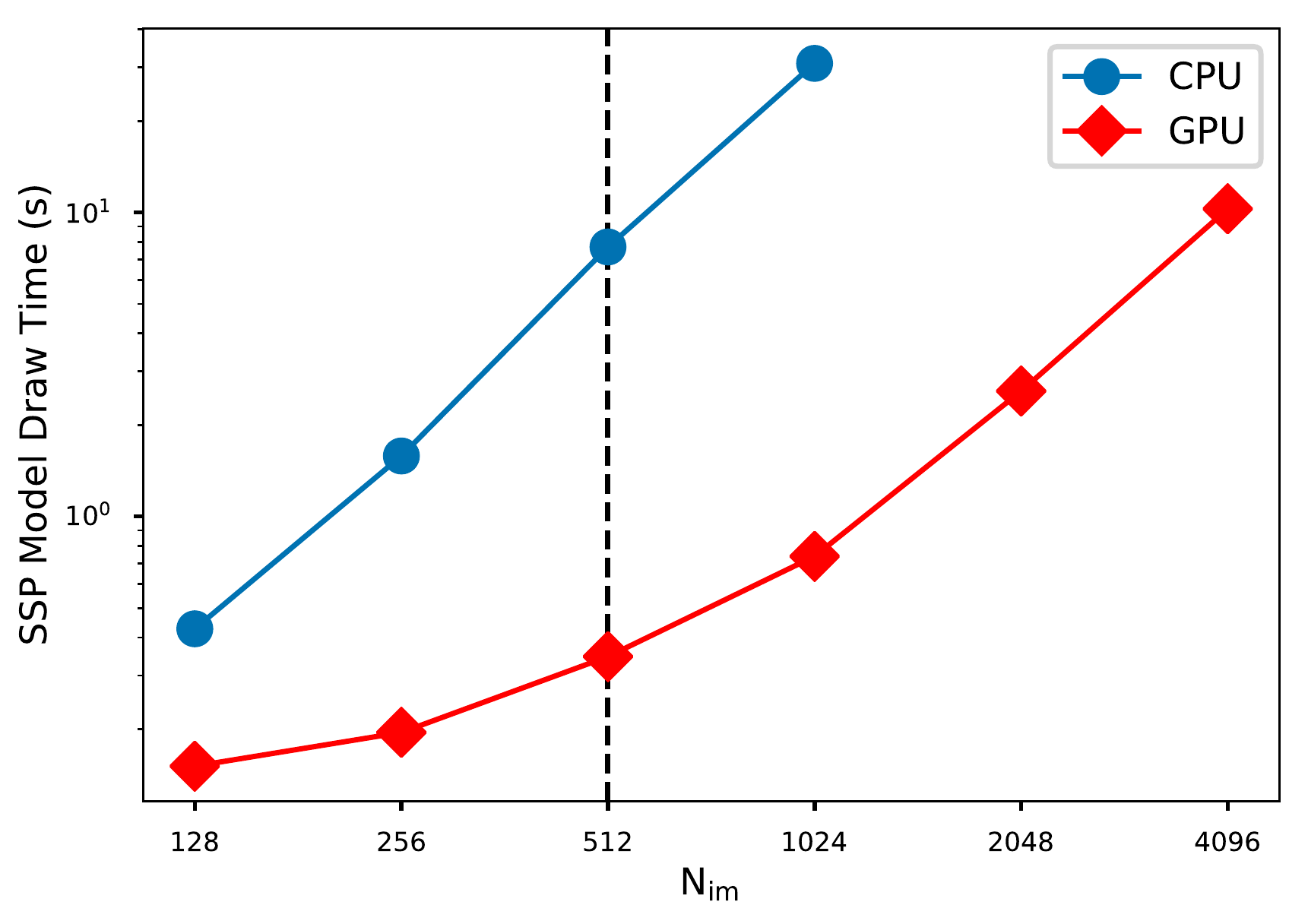}
    \caption{Computation time for drawing a SSP model pCMD is shown for the CPU and GPU-accelerated versions of the code, as a function of the simulated image size. For our fiducial model size of $\Nim = 512$, the GPU-accelerated code results in nearly $30\times$ speedup, and could be $4\times$ larger still with more modern GPU chips.}
    \label{f.gpu}
\end{figure}

Even when downsampling the isochrones by a factor of 5, simulating a pCMD  with the simplest SSP model requires $\mathcal{O}(10^8)$ random Poisson calls for a $512\times 512$ image.
Increasing the complexity and realism of a model by incorporating more age and metallicity points increases the computation time nearly linearly with the number of isochrones.
With 21 age points and $\sim5$ metallicity points, the more realistic physical models described in \S{ss.models} require $\mathcal{O}(10^{10})$ Poisson draws, and quickly become computationally infeasible on a traditional CPU.

\pcmdpy{} includes an optional \textit{GPU-accelerated} backend, which dramatically reduces the computational time required to simulate a pCMD.
Given the isochrones representing a particular physical model, each GPU thread independently samples the stars from those isochrones into an individual pixel.
This accelerates the simulation time of an individual model pCMD by a factor of $\sim 30\times$ compared to the CPU implementation. \F{f.gpu} shows the computation time required to simulate a pCMD of a single stellar population (model M1+S1) as a function of $\Nim$. The CPU tests were run on an Intel Xeon E5-2620 processor (2.1 GHz) and the GPU tests on an Nvidia Tesla K20Xm\footnote{These are the default GPUs available to us. The latest generation of Nvidia GPUs (the Tesla V100) could lead to an additional $\sim4\times$ speed-up.}.

The original code of \citet{Conroy2016} simulated an $\Nim=256$ pCMD for a relatively simple physical model (M1+S2+E1+D1) in $\approx 1$s.
The same computation in \pcmdpy{} takes only $\approx 0.25$s on an Nvidia Tesla K20Xm.
For our preferred model with a Gaussian MDF and 21 age bins (see \S{ss.mocks}), \pcmdpy{} simulates a pCMD with $\Nim=512$ in $\approx 2$ seconds.
GPU-acceleration allows for simulating both more complex (realistic) physical models and larger image sizes (less stochasticity in likelihoods, see \S{ss.sampling}) than would be feasible with CPUs alone.

\subsection{Likelihoods and Posterior Sampling}
\label{ss.sampling}

We evaluate the likelihood of a pCMD given a model with a binned Hess diagram and Gaussian statistics.
With this approach, we create a binned 2D histogram of pixels in the pCMD and compare the relative number of counts in the two distributions, normalized to the total number of pixels.
We choose bins of width $0.05$ magnitudes in each dimension, a width chosen to be roughly equivalent to the observational uncertainties, but we show in \S{ss.results} that the resulting posteriors are fairly insensitive to this choice. 
We compute the log-likelihood $\mathcal{L}$ using the counts of data pixels $d_i$ and model pixels $m_i$ in each Hess bin, as:
\begin{align}
    \mathcal{L} &\propto \sum_i \left[ -\frac{(\frac{d_i}{N_d} - \frac{m_i}{N_m})^2}{2 \sigma_i^2}\right] + \mathcal{C}\,,
\end{align}
where the uncertainty $\sigma_i$ is approximated by the square-root of the number of Hess bin counts, added in quadrature:
\begin{align}
    \sigma_i &= \max\left(\sqrt{d_i^2 + m_i^2}, 2\right)\, .
\end{align}
We apply a floor to the uncertainty to down-weight very rare bins in the Hess diagram, where differences in simulated and data image sizes may unintentionally bias the likelihood.
This is one aspect of a general problem of evaluating likelihoods in Hess diagram space: how to handle data points when there are no (or very few) model points.
For instance, a Poisson likelihood model would return zero likelihood if the model predicts zero pixels in a bin with even a single data pixels.

We add an additional likelihood term, corresponding to the difference in mean color $C$ and magnitude $M$ between the two distributions, with error of $0.05$ magnitudes:
\begin{align}
    \mathcal{C} &= -\frac{(C_d - C_m)^2}{2(0.05)^2} - \frac{(M_d - M_m)^2}{2(0.05)^2}.
\end{align}
Without this term, two pCMDs which do not overlap at all ($d_i = 0\;\forall\,i\; \mathrm{s.t.}\,m_i \neq 0$) are equally poor fits regardless of whether they are offset by an average of 1 or 10 magnitudes.
The addition of this term gives slight preference to models where the center of the distributions roughly align with the data, without substantially affecting best-fit estimates since the relative magnitude of the term is quite small.

We sample the posterior using a new python package for dynamic nested sampling, \dynesty{} (Speagle in prep). Nested sampling \citeeg{Skilling2004b, Feroz2009, Feroz2013, Handley2015} is an approach similar to the commonly-used Markov Chain Monte Carlo (MCMC) technique, representing the posterior through a collection of samples from the distribution.
Unlike MCMC, nested sampling efficiently computes the Bayesian evidence (also called marginal likelihood), allowing for principled model comparisons. Nested sampling algorithms also allow for sophisticated handling of multi-modal distributions. See the references above for more details on nested sampling.

Throughout this work, parameter estimates are reported as the median of the marginalized posterior probability function, and error-bars are reported as the $68\%$, equal-tailed credible interval, unless otherwise stated.

The pCMD likelihoods are stochastic, meaning that recalculating the likelihood multiple times for the same input parameters will result in a different log-likelihood.
This presents a statistical challenge for any posterior sampling algorithm, and without proper accounting leads to an underestimate in uncertainties \citep[see, however, the pseudo-marginal MCMC approach][]{Andrieu2009a}.
We discuss this further in Appendix \ref{a.likelihood}, and detail a method for post-processing the results to recover reasonable posteriors.

Fitting a pCMD with models of size $\Nim=512$ takes $\sim$100 GPU-hours for most cases.
This is a speed-up of a factor of $7\times$ compared to the original code of \citet{Conroy2016}, which required around 700 CPU-hours to fit a posterior using $\Nim=256$.
This was largely due to the need to combine the posteriors of 10 independent MCMC runs to overcome the stochasticity of the likelihoods.
The larger simulated image sizes allowed by GPU-acceleration decrease the inherent stochasticity of the likelihoods, making combining multiple fits unnecessary.

\section{A Suite of Mock Tests}
\label{s.tests}

\begin{figure*}
\plotonebig{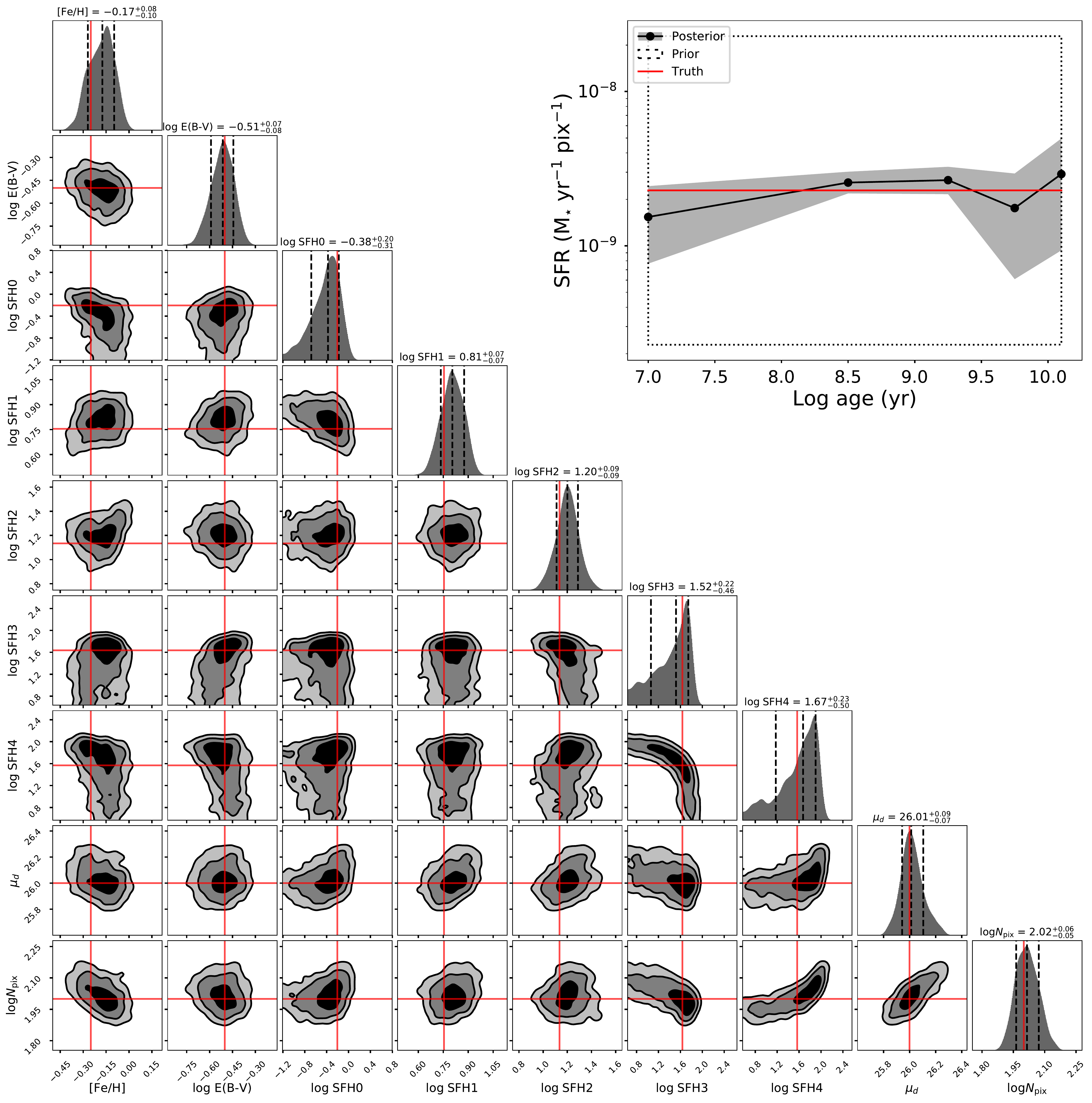}
\caption{Recovered posterior probability distribution from a mock test, using a 5-bin Non-Parametric SFH model to fit a constant star-formation rate model.
$\log\,\Npix$ is a derived quantity from the 5 SFH bins.
The $1\sigma$, $2\sigma$, and $3\sigma$ contours are shown.
The upper-right panel shows the recovered posterior estimates of star-formation rates as a function of time.
The true values used to generate the mock pCMD are shown with red lines.
The full model is specified in \S{ss.mocks} (fiducial-nonparametric model).
Every input parameter is recovered within the $68\%$ credible interval.}
\label{f.corner}
\end{figure*}

\begin{figure*}
\plotonebig{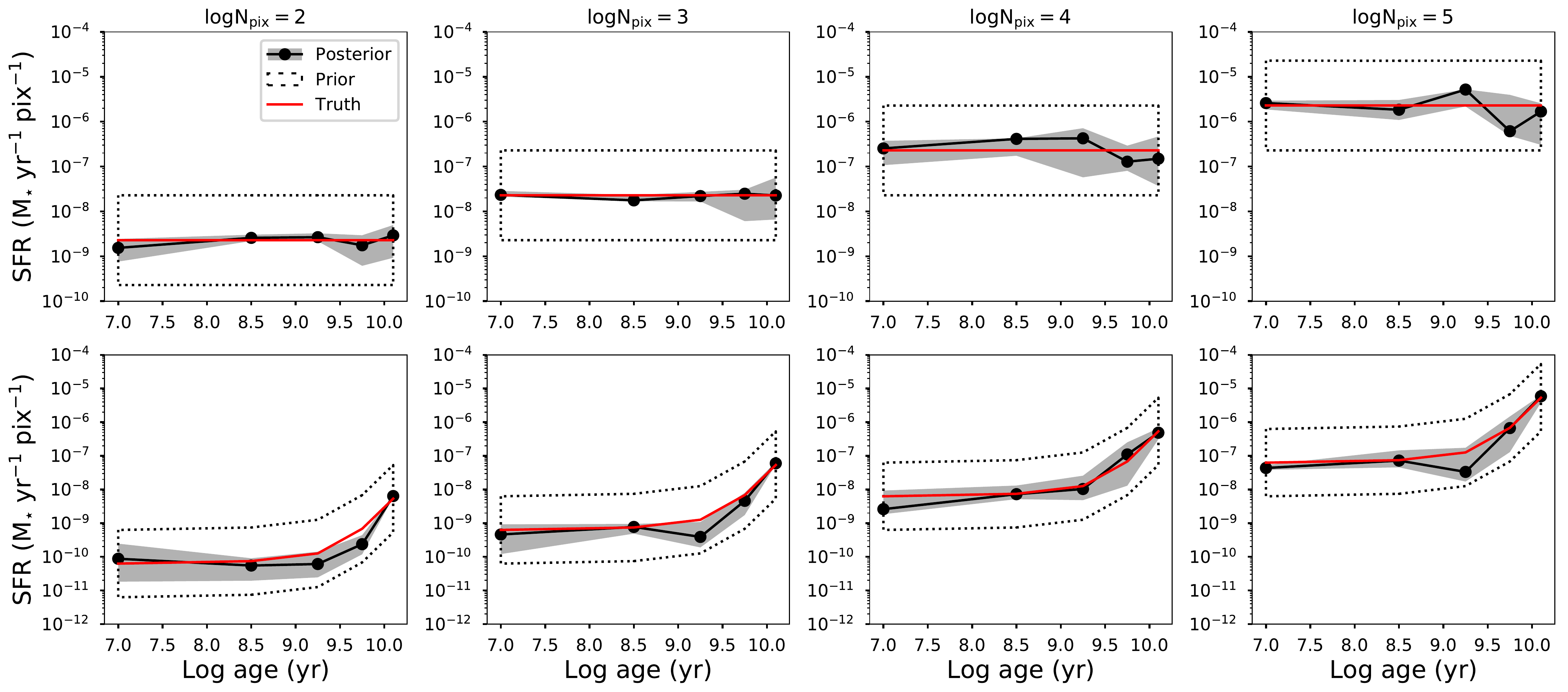}
\caption{Recovered SFH for several input models, as a function of \Npix{}. The input SFH is shown in red, while the posterior estimates (median and $68\%$ credible interval) are shown in grey. The prior, shown as a dotted outline, is assumed flat within $\pm 1$ dex of the true SFH. \textit{Top}: The input SFH has a constant star-formation rate. \textit{Bottom}: A tau-SFH model with $\tau=3$ Gyr.}
\label{f.SFH}
\end{figure*}

\begin{figure*}
\plotonebig{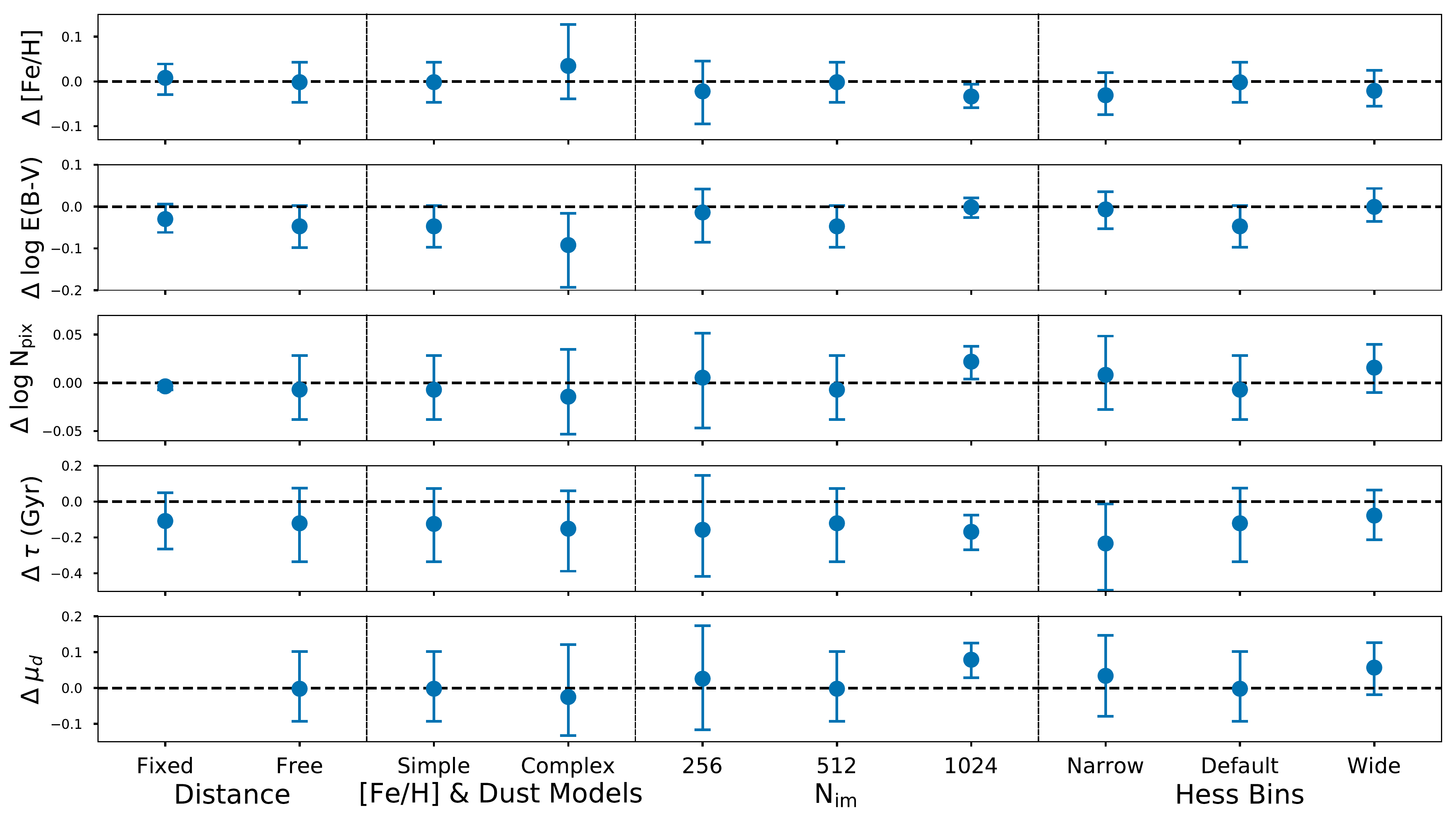}
\caption{Error in best fit (posterior median) and $68\%$ credible interval in each parameter for several mock tests with various modeling choices. The baseline model assumed is specified in \S{ss.mocks} (fiducial-$\tau$). \textit{Column 1}: Distance is held fixed in the first model. \textit{Column 2}: The simple models correspond to M2+E2, while the complex models correspond to M3+E3 (the widths \sigfeh{} and \sigdust{} are fit as free parameters). \textit{Column 3}: The size of the model image, $\Nim$. \textit{Column 4}: The width of the Likelihood Hess bins are either 0.02 dex (Narrow), 0.05 dex (Default), or 0.1 dex (Wide).}
\label{f.choices}
\end{figure*}

\subsection{Mock pCMD Models}
\label{ss.mocks}
To evaluate the capability of the code to recover model parameters, we run a series of mock tests where we fit models to simulated pCMDs generated from the code.
In most cases, we fit models with the same physical components (i.e.~same metallicity, SFH, and dust model) and hyper-parameters as used to generate the data.
We model observations in the F814W ($\mathrm{I}_{814}$) and F475W ($\mathrm{g}_{475}$) bands of ACS, use a mock image size of $\Nim=256$ and model image size of $\Nim=512$, and assume no sky noise.
Exposure times for F814W and F475W are set to 3235 and 3620 seconds, respectively, corresponding to the typical values of data from the PHAT survey \citep{Dalcanton2012}.

The fiducial model we frequently study, which we denote the "fiducial-$\tau$" model, has a fixed-width Gaussian MDF, tau-SFH, fixed-width log-normal dust screen, and the distance is allowed to vary (components M2+S3+E2+D2, see \S{ss.models} for details).
Unless otherwise specified, the free parameters used to generate the mock pCMDs are set to $\mufeh = -0.25$, $\mudust = -0.5$, $\Npix = 10^2$, $\tau = 3$ Gyr, and $\dmod=26.0$.
The width of the MDF is set to $\sigfeh = 0.2$, and the width of the Log-Normal dust model is set to $\sigdust = 0.1$. 

We also study a "fiducial-nonparametric" model, where the SFH is fit with a 5-bin, non-parametric model (S5).
The five SFH bins correspond to the following ages:
\begin{itemize}
    \item SFH0: 1 Myr $-$ 100 Myr
    \item SFH1: 100 Myr $-$ 1 Gyr
    \item SFH2: 1 Gyr $-$ 3 Gyr
    \item SFH3: 3 Gyr $-$ 10 Gyr
    \item SFH4: 10 Gyr $-$ 14 Gyr
\end{itemize}
Unless otherwise specified, the same free parameters are used to generate mock pCMDs as above, with the exception of the SFH, which is a constant star-formation rate model with $\Npix = 10^2$.

In both models, we assume flat priors over all parameters.
In most cases, we assume $\FeH \in [-0.5, +0.25]$, $\logEBV \in [-1, 0]$, $\log \Npix \in [2, 5]$, $\tau \in [0.1, 8.0]$ Gyr, and $\dmod \in [22, 26]$.
In the case of the non-parametric SFH, we assume flat priors in $\log \mathrm{SFH}$, of width $\pm 1$ dex around the true underlying SFH.

\subsection{Recovery of Non-Parametric SFHs}
\label{ss.results}

\begin{figure}
\plotonebig{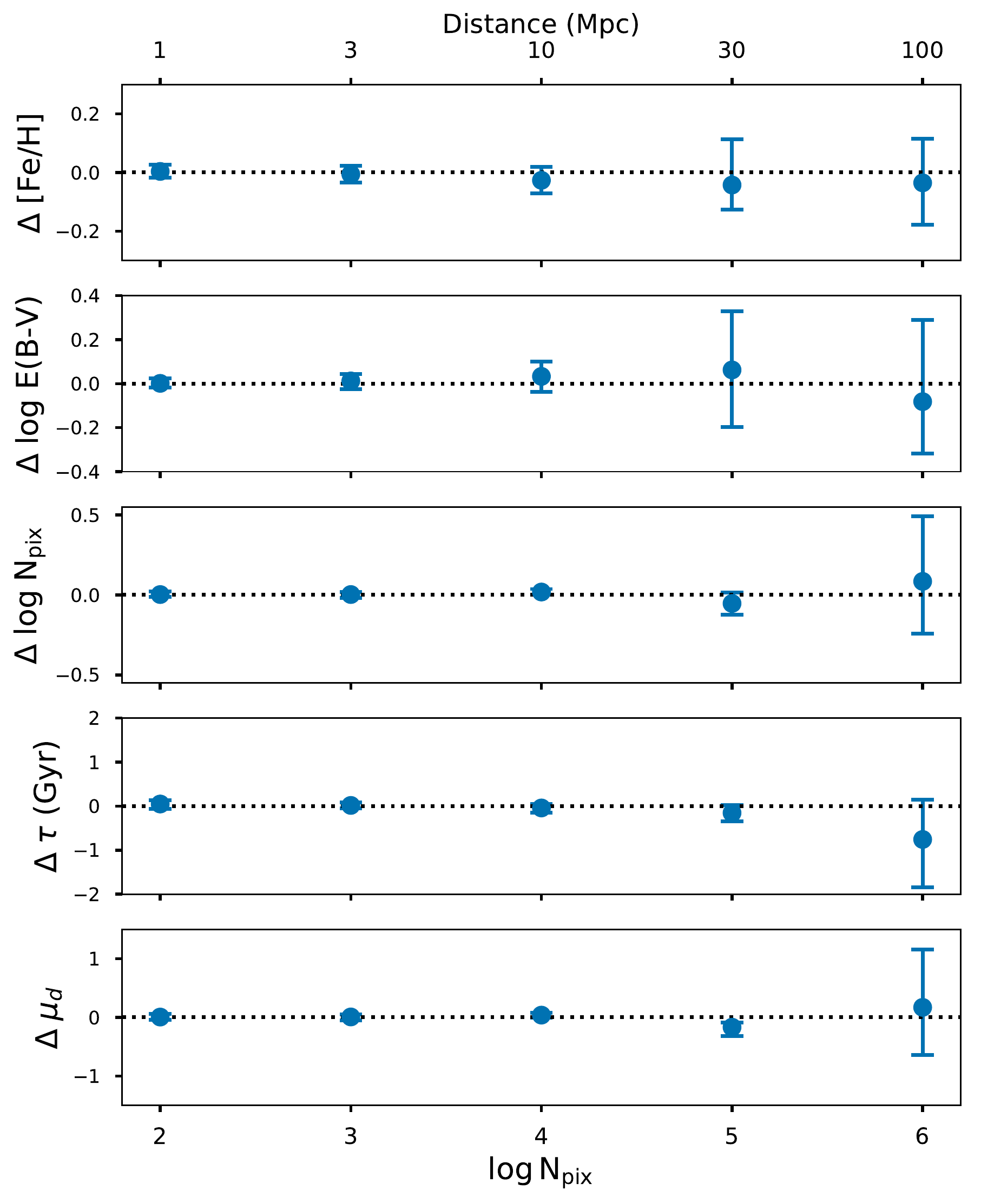}
\caption{Same as \F{f.choices}, where the same physical system is modeled at increasing distance. \Npix{} increases with distance, while \Nim{} of the mock pCMD is decreased to keep the simulated mass enclosed ($\approx 10^8 \mathrm{M}_\star$) constant. The model recovers highly precise estimates of all parameters out to $\approx 10$ Mpc, but the true results remain within the errors at distances as large as 100 Mpc.}
\label{f.Npix}
\end{figure}

\begin{figure}
\plotonebig{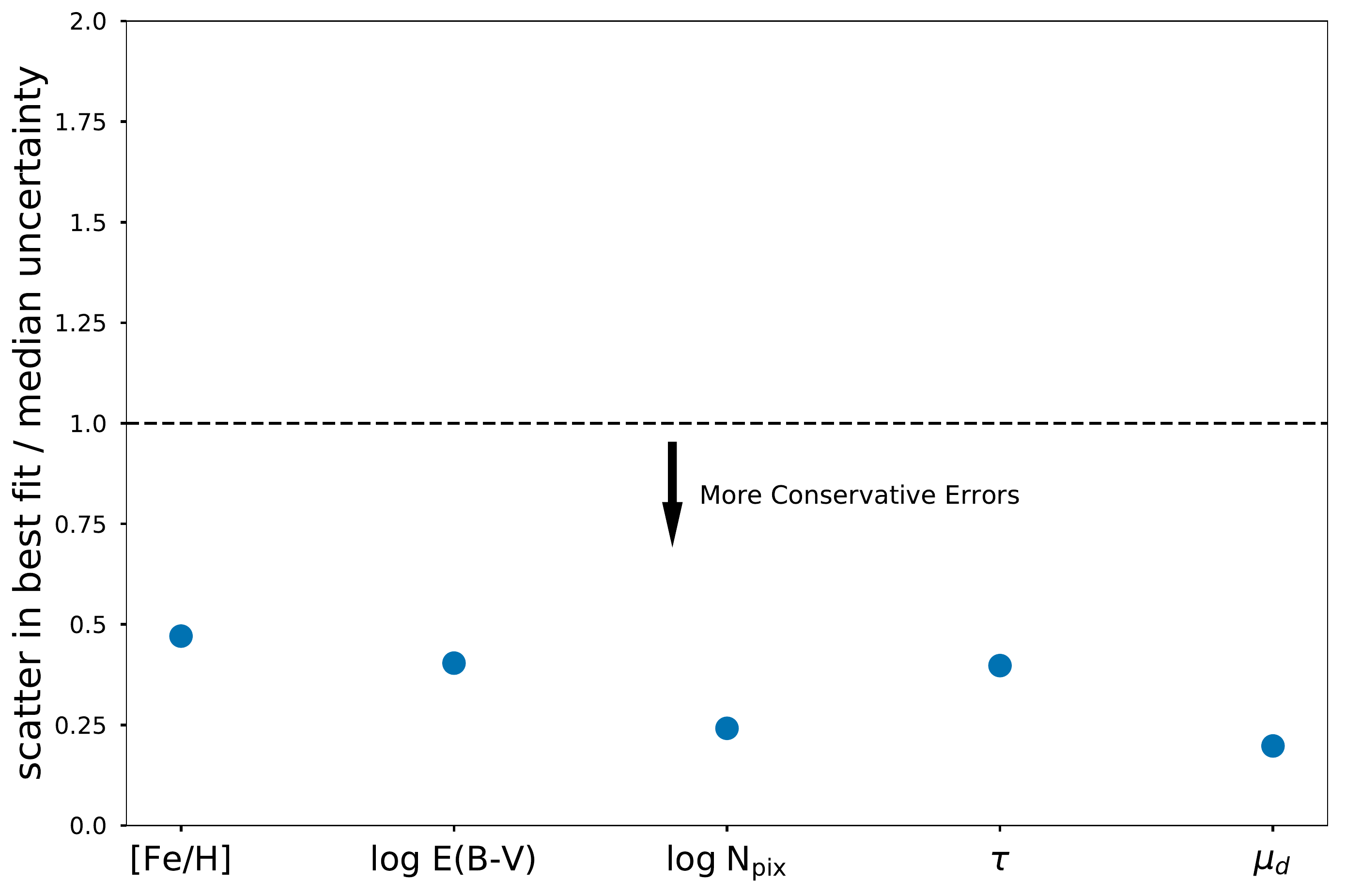}
\caption{The scatter in posterior median estimates of each model parameter is shown for 8 mock tests, evaluated on different realizations of the input model and normalized by the median uncertainty in each parameter. The fact that all values are below $1.0$ implies slightly overestimated uncertainties.}
\label{f.errors}
\end{figure}

\begin{figure*}
\plotonebig{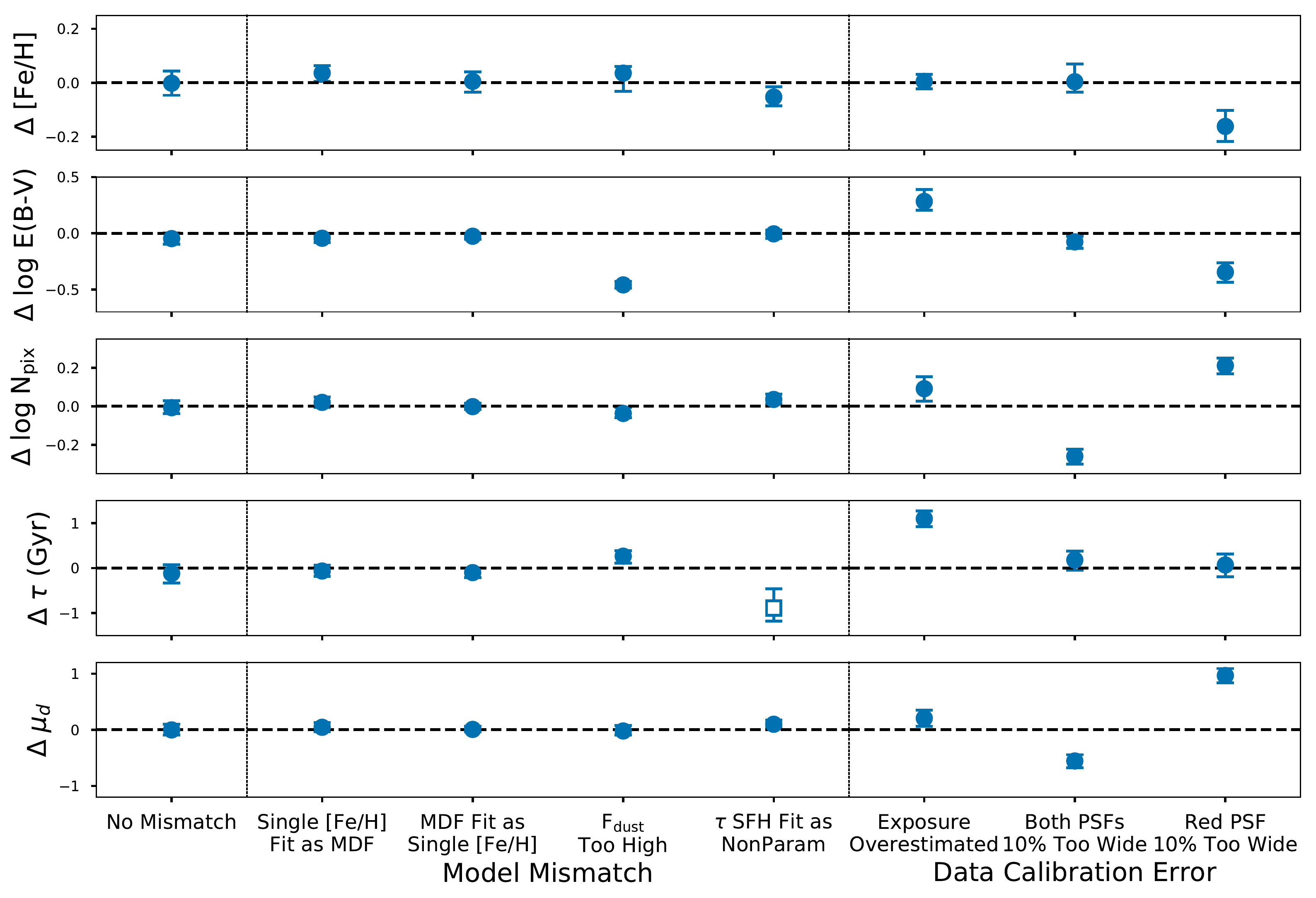}
\caption{Same as \F{f.choices}, for mock tests where there is a mismatch between the models used to generate the mock pCMD and to fit it.
\textit{Column 1}: There is no model mismatch.
\textit{Column 2}: Cases where the physical model assumed is incorrect. The white box shows the approximate $\tau$ derived from the non-parametric SFH.
\textit{Column 3}: Cases where observational hyper-parameters are miscalibrated.}
\label{f.mismatch}
\end{figure*}

\F{f.corner} shows the posterior probability distribution for a fiducial-nonparametric model fit, and demonstrates that \pcmdpy{} can simultaneously recover the input model parameters well.
The metallicity, dust content, distance, and total \Npix{} (computed as the sum of all SFH bins) are all recovered to within $1\sigma$, as shown in the marginalized histograms.
There is a notable degeneracy between \Npix{} and \dmod{}, but the distance modulus is still constrained to within 0.1 dex. 

The oldest bins of SFH (bins 3 and 4) are somewhat degenerate, but the total star formation older than 3 Gyr is fairly well constrained, as shown in the marginalized contours of those two bins.
\F{f.corner} also shows the derived constraints on the SFR in each pixel, as a function of age.
Compared to the original prior bounds allowed ($\pm 1$ dex in each bin), the model has recovered reasonably tight constraints on the SFH.

We also examine the recovered SFH for various input SFHs.
The results are shown in \F{f.SFH}, for constant-SFR and $\tau=3$ Gyr models and increasing \Npix{} from $10^2$ to $10^5$.
Each model is able to recover the underlying SFH, with the true SFH contained within the $68\%$ credible interval in nearly all cases.
In constant-SFR models, the oldest ages of star formation have the largest uncertainties, while the relatively higher SFR in the $\tau$ models is easier to constrain precisely.

\subsection{Evaluation of Model Choices}
\label{ss.evaluation}

We use the suite of mock tests to evaluate the effect of various hyper-parameter choices and model families on the recovered parameters.
For simplicity, these tests are performed using the fiducial-$\tau$ model, and the results are shown in \F{f.choices}.
These findings may not generalize fully to all models or regions of parameter space (especially higher \Npix), and we caution all users of \pcmdpy{} to think carefully about all hyper-parameter choices before fitting.

The first column of \F{f.choices} shows the effect of fitting distance as a free parameter (D1 vs.~D2).
We find we can recover the distance modulus to $\pm 0.1$ magnitudes.
With the exception of \Npix{} (which is slightly degenerate with \dmod{}), estimates of other parameters are just as well constrained regardless of whether distance is included in the fit.

The second column shows the effect of fitting for the width of the MDF and Dust distributions (Simple: M2+E2. Complex: M3+E3).
We find the model is unable to constrain these widths, which leads to a small bias and inflated uncertainties in the estimates of other parameters.
We therefore recommend against fitting for these width parameters, unless available information suggests a fairly informative prior is warranted.

In the third column, we vary the model image size \Nim{}. Smaller image sizes ($\Nim = 256$) lead to inflated uncertainties, despite taking just as long to fit as the default size of $\Nim = 512$.
The additional stochasticity in the likelihoods at small \Nim{} leads to more severe drops in sampling efficiency (see Appendix \ref{a.likelihood}).
Larger image sizes ($\Nim = 1024$) take significantly longer to fit, requiring fewer nested sampling live points to converge in a reasonable time (fewer than 300-GPU hours), which results in under-estimated errors.

The final column shows various choices for the width of the Hess bins used to evaluate likelihoods.
The bin size has no significant effect on the recovery of the posterior, although we note that this may not hold true at higher \Npix{}, when the pCMD distribution is more compact.

In \F{f.Npix}, we study the ability of \pcmdpy{} to model galaxies as a function of distance.
At larger distances, the magnitude of surface-brightness fluctuations decreases (since \Npix{} is larger), making the models less sensitive to the underlying stellar populations.
In addition, the same physical size region will correspond to far fewer pixels, and gradients in surface-brightness and stellar populations will make the modeling assumptions invalid over too many pixels.

To simulate these effects, we begin with a fiducial-$\tau$ model with $\Npix = 10^2$, modeled at distance of 1 Mpc, roughly equivalent to the disk of a galaxy like M31 \citep[][see also Cook et al. In Prep]{Conroy2016}.
The mock pCMD for this system is generated with $\Nim = 2048$, corresponding to a region 500pc on a side with total mass $\mathrm{M}_\star \approx 10^{8}\,\mathrm{M}_\odot$.
We then approximate the effects of observing the same physical system at larger distance and correspondingly-higher \Npix{}, while decreasing \Nim{} of the mock pCMD to keep the physical size and total mass constant.
Five such mock datasets (from 1 to 100 Mpc) are fit to the same model (with simulated $\Nim = 512$). 

The fits from this experiment are in excellent agreement with the true models until $\mathrm{D} \gsim 10$ Mpc ($\Npix \gsim 10^4$), at which point the uncertainty rises sharply.
But even out to 100 Mpc, the true parameters remain within the $68\%$ credible interval. 

The pCMD method should therefore provide interesting constraints of stellar populations and distance out to large distances, and may serve as a useful complement to the existing SED-modeling technique.

To test the stability of the results, we run eight fits on different realizations of pCMDs generated with the same underlying model.
The scatter in the estimates of each fit should fall within the typical uncertainty, or else the uncertainties are likely underestimated.
\F{f.errors} shows the results of this test.
For all parameters, the scatter (standard-deviation) in median estimates is significantly less than the typical uncertainty.
This suggests that the model uncertainties are likely \textit{over-estimated} by a factor of roughly $2\times$. 
It could be possible to correct for this with a more careful selection of $\Lmax$, defined in Appendix \ref{a.likelihood}, but we are comfortable with over-estimating our uncertainties, given the relatively loose method for dealing with the stochastic likelihoods detailed in the Appendix.

\subsection{Model Mismatch Tests}
\label{ss.mismatch}

We investigate the effect of fitting a pCMD with a different model than used to generate the data in \F{f.mismatch}.
First, we show fits where the physical model is incorrect.
Incorrectly modeling the metallicity distribution function may lead to a minor systematic bias in \FeH{} and $\tau$, but for $\sigfeh{}$ of order a few-tenths of a dex, the effects are small and within the uncertainties.

Slight biases also arise from incorrectly modeling the SFH or the structure of the dust. 
We fit a model with $\Fdust = 1.0$ to data generated with $\Fdust = 0.5$, effectively ignoring that half of the stars should lie above the screen of dust.
This results in an underestimate of the dust content, as well as bias in $\tau$ and \Npix{}.
Accurately accounting for the geometries of stars relative to the dust is therefore important to recovering accurate measurements.

We also show the effect of modeling a $\tau$-SFH with a non-parametric model.
For comparison to the $\tau$ models, we compute the average age of the inferred SFHs, and convert to an effective $\tau$.
The non-parametric model appears slightly biased towards preferring old ages (lower $\tau$).
This results in slight positive bias in \Npix{} and \dmod{} in order to result in the same total luminosity.
The age bias could be mitigated by increasing the number of SFH bins modeled, but this becomes computationally demanding.
Incorrectly specifying the physical model can lead to subtle systematic biases, and it is not always easy to diagnose such a mismatch.
One possible approach could be to compare the Bayesian evidence (computed through the nested sampling algorithm) for multiple model choices, or alternatively by studying the residual patterns in Hess diagram space.
In future work, we additionally intend to study the effect of different SFH priors, such as those suggested by \citet{Leja2018b}.

Other possible mismatches between data and model include errors in the calibration of the observational data.
If the exposure time of images is overestimated (the model shown overestimates the true exposure by $2\times$), the model greatly overpredicts the age parameter $\tau$, and most other parameters are also biased from their true values.
When extracting pCMDs from observations, it is important to properly understand the observing conditions and whether multiple exposures were co-added to produce the final photometry.

The final two examples showcase the complex effects of poor PSF calibrations.
If both PSF models are too wide (FWHM overestimated by $10\%$), best fit models are strongly biased in all parameters.
Yet if only one PSF model is miscalibrated (here, the F814W filter), many of the biases go in the opposite direction.
We investigate this effect more fully in upcoming work (Cook et al., In Prep) when we discuss the important practical challenges in selecting observational data to analyze with the pCMD technique.

\section{Conclusions}
\label{s.conclusions}

We have presented a study of the application of pixel color-magnitude diagrams (pCMDs) for studying galaxies in the semi-resolved regime, as first introduced by \citet{Conroy2016}.
We developed the open-source, GPU-accelerated Python package \pcmdpy{}, detailed its primary modeling assumptions and physical models implemented.
We also presented the results of a suite of mock tests that demonstrate the potential for using \pcmdpy{} to constrain important physical parameters of galaxies.

The main results of this work are as follows:
\begin{enumerate}
    \item We have developed the new package \pcmdpy{} for modeling pCMDs and inferring the physical properties of galaxies. It contains a number of improved physical models compared to the original code of \citet{Conroy2016}, including a Gaussian metallicity-distribution function, a Log-Normal dust extinction model, and allows for simultaneous fitting of distance modulus. The code uses GPU-acceleration to allow the modeling of pCMDs with these complex models and larger simulated image sizes in significantly less time. Posteriors are sampled using the dynamic nested sampling code \dynesty{}, and fitting a model requires $\sim 1/7^\mathrm{th}$ as much computational time as the original code.
    
    \item We demonstrated that pCMDs are sensitive to the distance to a galaxy, and that our code can simultaneously recover accurate estimates of distance, SFH, metallicity, and dust content.
    
    \item We show that highly precise measurements of distance and stellar populations should be recoverable for massive galaxies at least to 10 Mpc with \HST{}-like resolution, and reasonable constraints are possible out to 100 Mpc.
    
    \item The model is relatively robust against small biases in the assumed metallicity distribution function. Mischaracterized dust or SFH models can lead to small systematic bias in recovered parameters. Additionally, it is crucial to have good estimates of optical properties such as exposure time and well-calibrated PSF models. Future work will discuss other important considerations for selecting observational data to fit with \pcmdpy{}.
    
    \item The effects of the stochastic likelihood model present challenges for traditional sampling methods. The larger images that can be simulated using GPUs reduce this burden somewhat, but additional steps must be taken to account for the bias towards positive-likelihood fluctuations. Our chosen approach may result in overestimating the derived uncertainties by a factor of order $2\times$.
\end{enumerate}

\acknowledgments
We thank Aaron Dotter and Jieun Choi for helpful discussions around the details of the MIST models. B.C.~acknowledges support from the NSF Graduate Research Fellowship Program under grant DGE-1144152. This work is supported in part by HST-AR-14557. The computations in this paper were run on the Odyssey cluster supported by the FAS Division of Science, Research Computing Group at Harvard University.

\software{This research has made use of NASA's Astrophysics Data System, as well as the following software packages: PyCUDA \citep{Klockner2012}, Dynesty (Speagle in prep), NumPy \citep{VanderWalt2011}, Matplotlib \citep{Hunter2007}, IPython \citep{Perez2007}, Jupyter \citep{Kluyver2016}, SciPy \citep{Jones2001},
Pandas \citep{McKinney2010}, and Astropy \citep{TheAstropyCollaboration2013,TheAstropyCollaboration2018}.}

\appendix
\section{Stochastic Likelihoods}
\label{a.likelihood}

The forward-modelling procedure for generating pCMDs is stochastic, and therefore so are our likelihoods.
The simulated pCMD is a random realization of the underlying model, and therefore evaluating the likelihood of the same point in parameter-space multiple times will produce different results.
Our tests indicate the distribution of likelihoods, given fixed model parameters, has a well-defined center but a long positive tail, essentially resulting from "overfitting" to the noise in the data.
Larger simulated image sizes $\Nim$ produce pCMDs which average over the many rare fluctuations, and therefore decrease the inherent stochasticity.

Any level of stochasticity poses significant problems for nested sampling or MCMC, as a positive fluctuation in likelihood will result in a model point being given larger weight than it truly deserves.
In either MCMC or nested sampling, this has the detrimental effect of significantly decreasing sampling efficiency.
The sampling algorithms will take longer to find a comparably good fit, resulting in decreased sampling efficiency and a cascading effect where most subsequent sampled points are also drawn from the positive-fluctuation tails.

Our tests have identified two other detrimental effects of nested sampling in a stochastic likelihood model.
The weights assigned to the sampled points are biased towards only weighting the largest fluctuations, often resulting in only a single point being assigned a non-trivial weight.
Since nested sampling integrates towards increasing likelihood, the final many points returned by the algorithm, and those assigned highest weight, almost all represent positive likelihood fluctuations, rather than actually better models.

Furthermore, the auto-stopping criteria used by many algorithms, including \dynesty{}, relies on $\Delta \ln Z$, the estimated evidence remaining \citeeg{Feroz2009}.
This value is biased high by the positive fluctuations, leading the algorithm to continue sampling longer than would be desired.
Combined with the decrease in sampling efficiency, our \dynesty{} fits often become stuck, taking thousands of likelihood calls to return a new sample\footnote{An example of this principle, for a simplified case of a stochastic likelihood, is included as a demonstration in the Dynesty respository:\url{https://github.com/joshspeagle/dynesty/blob/master/demos/Examples -- Noisy Likelihoods.ipynb}}.
In practice, we stop the model fit after a fixed number of iterations.
As discussed below, we find that we can later account for the bias in likelihoods, and when we do so the fits are almost always well converged by the $\Delta \ln Z$ criterion.

\begin{figure}
\plotoneman{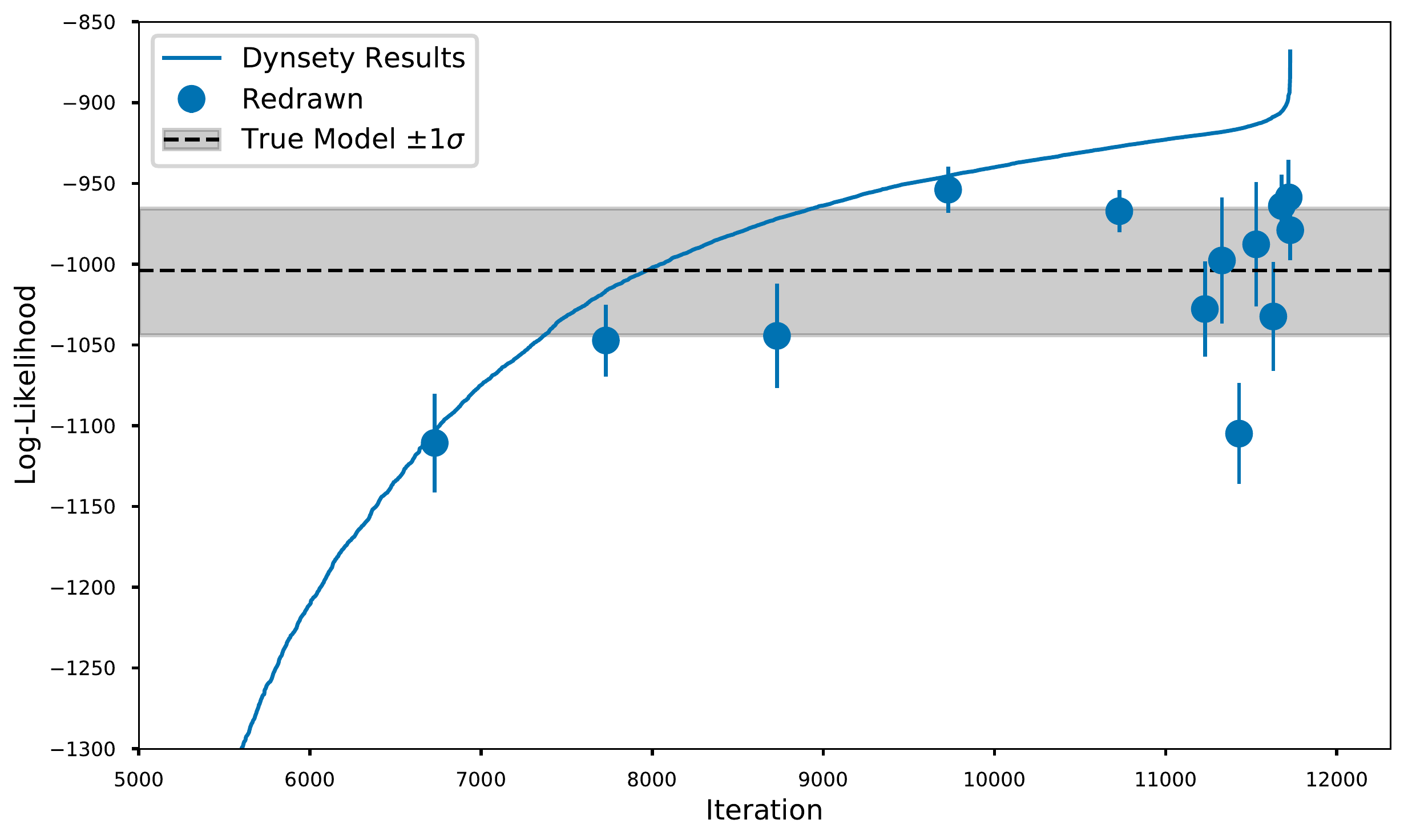}{0.8}
\caption{A demonstration of the bias induced from sampling from a stochastic likelihood function. The blue line shows the log-likelihood reported for each nested-sampling point of a mock test as a function of iteration. The blue error-bars show the distribution of log-likelihood from recomputing the log-likelihood of the same points in parameter space 10 times. The distribution of log-likelihoods computed using the true (known) model parameters is shown in black, and any log-likelihood above that range is simply a rare positive-fluctuation that is over-fitting the stochastic effects.}
\label{f.resampling}
\end{figure}

We show the effect of this likelihood bias in \F{f.resampling}.
The blue line shows the measured log-likelihood of the \dynesty{} samples from a mock test. The upturn at the final points shows extreme positive fluctuations: using the results straight from \dynesty{}, the final point has $99\%$ of the weight, producing an implausibly peaked posterior.
To demonstrate these points all represent positive likelihood fluctuations, we recomputed the likelihood of several of the points in parameter space, and show the measured distribution of likelihoods as error-bars.

It is clear that the likelihoods returned by \dynesty{} are biased high from the average likelihood for each model.
We can estimate what the maximum realistic likelihood should be, by computing many realizations of the "true" model used to generate the pCMD, and evaluating the fits.
This is shown in the grey band.
As should be expected, all of the mean likelihoods for the resampled points lie roughly within this band.
We therefore have a means for estimating which likelihoods are trustworthy, and which are likely biased high due to fluctuations.

We use this insight to adopt a \textit{post-processing procedure} to account for this likelihood bias.
At the completion of each \dynesty{} run, we recompute the likelihood of the best fit model 100 times, and then adopt the median value as a likelihood threshold, \Lmax.
Any likelihoods above that threshold are capped to that value, and we recompute the weights and convergence statistics using the updated likelihoods.

\begin{figure}
\plotoneman{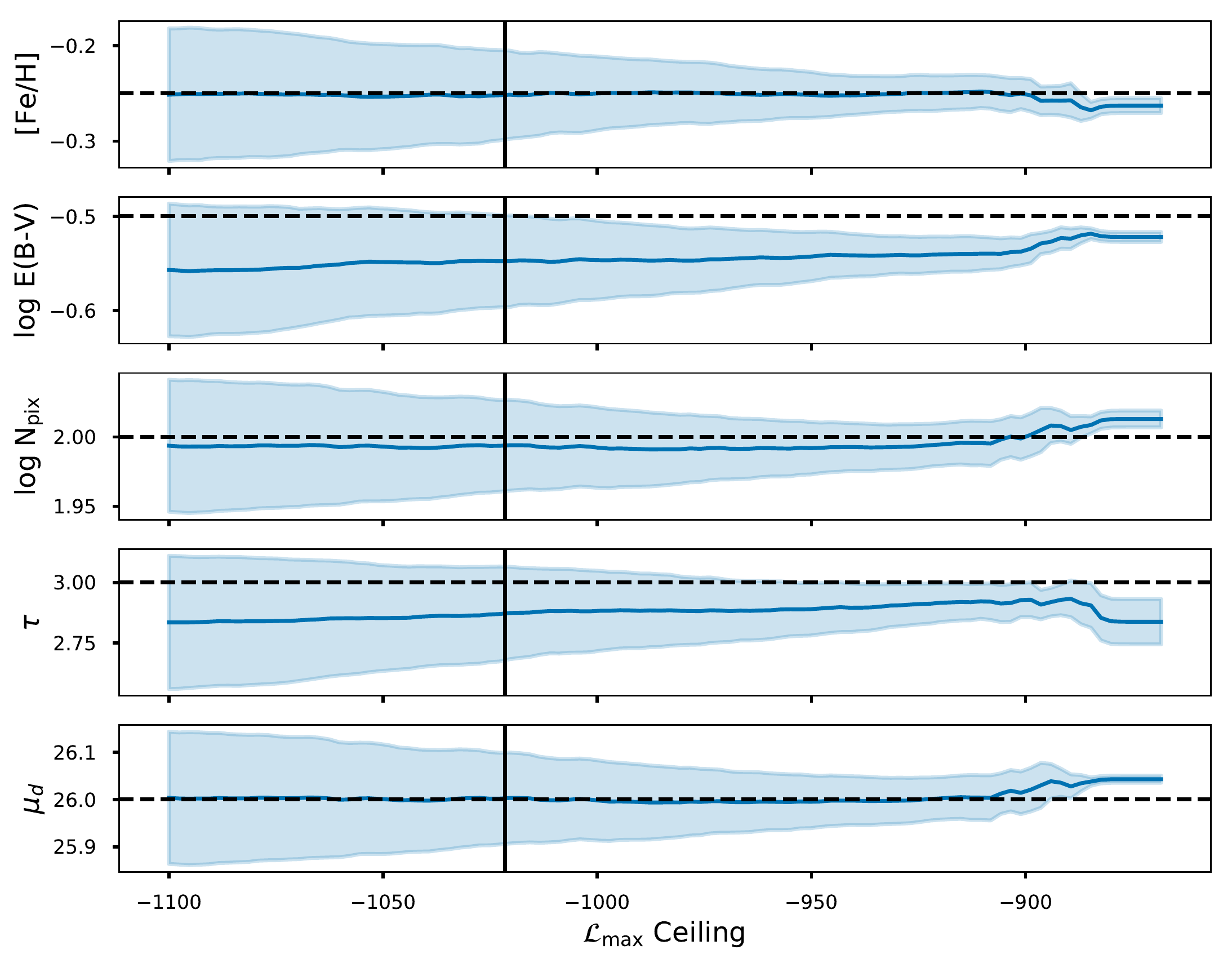}{0.8}
\caption{The marginalized mean estimates (solid line) and uncertanties in each parameter as a function of the maximum log-likelihood ceiling applied. The solid black line shows the adopted threshold: the median log-likelihood of the best-fit model recomputed $100\times$ against the data. The dashed lines show the true parameters.}
\label{f.ceiling}
\end{figure}

We find that this approach is successful at eliminating the bias from likelihood fluctuations, by down-weighting the final handful of samples that are most subject to this stochastic bias.
\F{f.ceiling} shows the mean and uncertainties in model parameters recovered as a function of this \Lmax.
When \Lmax{} is too high, the mean estimates vary sharply, as only one or a few sampled points are given all the posterior weight.
Using our adopted \Lmax{} ceiling, we are averaging over sufficiently many sampled points to recover stable means and reasonable uncertainties.
When the remaining-evidence criterion is recomputed using the adjusted likelihoods, the fits are almost always extremely well converged by standard stopping criteria ($\Delta \ln Z \lsim 0.01$).

This method for handling the stochastic likelihoods is a rough, heuristic approach, so the exact statistical uncertainties must be treated cautiously. The mock tests shown in this work are therefore important for validating that the approach recovers reasonable errors. As shown in \F{f.errors}, the scatter in mean estimates between multiple independent trials is smaller than the typical uncertainty, indicating that if anything this relatively ad-hoc procedure is somewhat overly conservative in the model uncertainties.

\section{Cloud Computing and GPUs}
\label{a.cloud}

The \pcmdpy{} GPU-acceleration is written in \texttt{CUDA}, a proprietary language that only operates on Nvidia-brand GPUs. This limits the systems that can utilize the GPU-accelerated version of \pcmdpy{}.
Most personal computers and laptops (especially Macs) do not have Nvidia cards; Intel and AMD are common competitors, but CUDA code cannot be compiled to run on their architecture.
However, the GPUs installed in most compute clusters hosted at research institutions are Nvidia, and cloud computing offers an alternative option for obtaining GPU-accelerated computational resources.

To make \pcmdpy{}'s GPU capabilities more broadly applicable, we have also developed a Docker container architecture for fitting pCMDs in the cloud using AWS-Batch.
Using this framework, pCMD models can be fit by renting GPU-enabled instances on-demand, with the benefit of having practically limitless scaling potential (large numbers of jobs can be run simultaneously).
As of this writing, such instances are available for around $\$ 1$/GPU-hour (with faster GPUs available for higher cost, in nearly linear proportion to their speed-up).
As \pcmdpy{} typically takes around 100 GPU-hours to fit a model, this translates to around $\$100$ per model fit.
Full details and examples will be detailed later with the public \pcmdpy{} code release paper.
For a detailed example on using AWS in astronomical research, see \citet{Williams2018a}.

\bibliography{references}
\end{document}

%% file: setup.tex
\usepackage[T1]{fontenc}
\usepackage{microtype}
\usepackage{afterpage,amsmath,amssymb,graphicx,natbib, color}
\usepackage{hyperref}
\usepackage[english]{babel}
\usepackage{blindtext}
\usepackage{array}   
\newcommand\plotonebig[1]{\centering \leavevmode
\includegraphics[width=.98\linewidth]{#1}}

\newcommand\plotoneman[2]{\centering \leavevmode
\includegraphics[width=#2\linewidth]{#1}}

\newcommand\citeeg[1]{\citep[e.g.,][]{#1}}
\newcommand\lsim{\lesssim}
\newcommand\gsim{\gtrsim}

\newcommand\Npix{\ensuremath{\mathrm{N_{pix}}}}
\newcommand\Nage{\ensuremath{\mathrm{N_{age}}}}
\newcommand\Nim{\ensuremath{\mathrm{N}_\mathrm{im}}}
\newcommand\Nbands{\ensuremath{\mathrm{N}_\mathrm{bands}}}
\newcommand\FeH{\ensuremath{[\mathrm{Fe}/\mathrm{H}]}}
\newcommand\logEBV{\ensuremath{\log \mathrm{E}(\mathrm{B}-\mathrm{V})}}
\newcommand\dmod{\ensuremath{\mu_{d}}}
\newcommand\mufeh{\ensuremath{\FeH}}
\newcommand\sigfeh{\ensuremath{\sigma_{\FeH}}}
\newcommand\mudust{\ensuremath{\logEBV}}
\newcommand\sigdust{\ensuremath{\sigma_{\mathrm{dust}}}}
\newcommand\Fdust{\ensuremath{\mathrm{F}_\mathrm{dust}}}
\newcommand\Lmax{\ensuremath{\mathcal{L}_\mathrm{max}}}
\newcommand\changed[1]{{\bf #1}}

\newcommand\pcmdpy{\texttt{PCMDPy}}
\newcommand\dynesty{\texttt{dynesty}}
\renewcommand\S[1]{\text{Section }\ref{#1}}
\newcommand\F[1]{\text{Figure }\ref{#1}}
\newcommand\T[1]{\text{Table }\ref{#1}}

\newcommand\HST{\textit{HST}}

\renewcommand\changed[1]{ #1}

%% file: paper1.bbl
\begin{thebibliography}{48}
\expandafter\ifx\csname natexlab\endcsname\relax\def\natexlab#1{#1}\fi

\bibitem[{Andrieu \& Roberts(2009)}]{Andrieu2009a}
Andrieu, C., \& Roberts, G.~O. 2009,
  \href{http://dx.doi.org/10.1214/07-AOS574}{The Annals of Statistics, 37, 697}

\bibitem[{Cantiello {et~al.}(2018)Cantiello, Jensen, Blakeslee, Berger, Levan,
  Tanvir, Raimondo, Brocato, Alexander, Blanchard, Branchesi, Cano, Chornock,
  Covino, Cowperthwaite, D’Avanzo, Eftekhari, Fong, Fruchter, Grado, Hjorth,
  Holz, Lyman, Mandel, Margutti, Nicholl, Villar, \& Williams}]{Cantiello2018}
Cantiello, M., Jensen, J.~B., Blakeslee, J.~P., {et~al.} 2018,
  \href{http://dx.doi.org/10.3847/2041-8213/aaad64}{The Astrophysical Journal,
  854, L31}

\bibitem[{Carlsten {et~al.}(2019)Carlsten, Beaton, Greco, \&
  Greene}]{Carlsten2019a}
Carlsten, S., Beaton, R., Greco, J., \& Greene, J. 2019,
  \href{http://arxiv.org/abs/1901.07575}{eprint arXiv:1901.07575}

\bibitem[{Carnall {et~al.}(2018)Carnall, Leja, Johnson, McLure, Dunlop, \&
  Conroy}]{Carnall2018b}
Carnall, A.~C., Leja, J., Johnson, B.~D., {et~al.} 2018,
  \href{http://arxiv.org/abs/1811.03635}{eprint arXiv:1811.03635}

\bibitem[{Choi {et~al.}(2016)Choi, Dotter, Conroy, Cantiello, Paxton, \&
  Johnson}]{Choi2016}
Choi, J., Dotter, A., Conroy, C., {et~al.} 2016,
  \href{http://dx.doi.org/10.3847/0004-637X/823/2/102}{The Astrophysical
  Journal, 823, 102}

\bibitem[{Cohen {et~al.}(2018)Cohen, Dokkum, Danieli, Romanowsky, Abraham,
  Merritt, Zhang, Mowla, Kruijssen, Conroy, \& Wasserman}]{Cohen2018}
Cohen, Y., Dokkum, P.~v., Danieli, S., {et~al.} 2018,
  \href{http://dx.doi.org/10.3847/1538-4357/aae7c8}{The Astrophysical Journal,
  868, 96}

\bibitem[{Conroy(2013)}]{Conroy2013b}
Conroy, C. 2013,
  \href{http://dx.doi.org/10.1146/annurev-astro-082812-141017}{Annual Review of
  Astronomy and Astrophysics, 51, 393}

\bibitem[{Conroy \& van Dokkum(2016)}]{Conroy2016}
Conroy, C., \& van Dokkum, P.~G. 2016,
  \href{http://dx.doi.org/10.3847/0004-637X/827/1/9}{The Astrophysical Journal,
  827, 9}

\bibitem[{Conroy {et~al.}(2015)Conroy, van Dokkum, \& Choi}]{Conroy2015}
Conroy, C., van Dokkum, P.~G., \& Choi, J. 2015,
  \href{http://dx.doi.org/10.1038/nature15731}{Nature, 527, 488}

\bibitem[{Dalcanton {et~al.}(2012)Dalcanton, Williams, Lang, Lauer, Kalirai,
  Seth, Dolphin, Rosenfield, Weisz, Bell, Bianchi, Boyer, Caldwell, Dong,
  Dorman, Gilbert, Girardi, Gogarten, Gordon, Guhathakurta, Hodge, Holtzman,
  Johnson, Larsen, Lewis, Melbourne, Olsen, Rix, Rosema, Saha, Sarajedini,
  Skillman, \& Stanek}]{Dalcanton2012}
Dalcanton, J.~J., Williams, B.~F., Lang, D., {et~al.} 2012,
  \href{http://dx.doi.org/10.1088/0067-0049/200/2/18}{The Astrophysical Journal
  Supplement Series, 200, 18}

\bibitem[{Dalcanton {et~al.}(2015)Dalcanton, Fouesneau, Hogg, Lang, Leroy,
  Gordon, Sandstrom, Weisz, Williams, Bell, Dong, Gilbert, Gouliermis,
  Guhathakurta, Lauer, Schruba, Seth, \& Skillman}]{Dalcanton2015}
Dalcanton, J.~J., Fouesneau, M., Hogg, D.~W., {et~al.} 2015,
  \href{http://dx.doi.org/10.1088/0004-637X/814/1/3}{The Astrophysical Journal,
  814, 3}

\bibitem[{Dolphin(2002)}]{Dolphin2002}
Dolphin, A.~E. 2002,
  \href{http://dx.doi.org/10.1046/j.1365-8711.2002.05271.x}{Monthly Notices of
  the Royal Astronomical Society, 332, 91}

\bibitem[{Feroz {et~al.}(2009)Feroz, Hobson, \& Bridges}]{Feroz2009}
Feroz, F., Hobson, M.~P., \& Bridges, M. 2009,
  \href{http://dx.doi.org/10.1111/j.1365-2966.2009.14548.x}{Monthly Notices of
  the Royal Astronomical Society, 398, 1601}

\bibitem[{Feroz {et~al.}(2013)Feroz, Hobson, Cameron, \& Pettitt}]{Feroz2013}
Feroz, F., Hobson, M.~P., Cameron, E., \& Pettitt, A.~N. 2013,
  \href{http://arxiv.org/abs/1306.2144}{eprint arXiv:1306.2144}

\bibitem[{Greene {et~al.}(2019)Greene, Veale, Ma, Thomas, Quenneville,
  Blakeslee, Walsh, Goulding, \& Ito}]{Greene2019}
Greene, J.~E., Veale, M., Ma, C.-P., {et~al.} 2019,
  \href{http://arxiv.org/abs/1901.01271}{eprint arXiv:1901.01271}

\bibitem[{Handley {et~al.}(2015)Handley, Hobson, \& Lasenby}]{Handley2015}
Handley, W.~J., Hobson, M.~P., \& Lasenby, A.~N. 2015,
  \href{http://dx.doi.org/10.1093/mnras/stv1911}{Monthly Notices of the Royal
  Astronomical Society, 453, 4384}

\bibitem[{Hopkins {et~al.}(2014)Hopkins, Kere{\v{s}}, O{\~{n}}orbe,
  Faucher-Gigu{\`{e}}re, Quataert, Murray, \& Bullock}]{Hopkins2014}
Hopkins, P.~F., Kere{\v{s}}, D., O{\~{n}}orbe, J., {et~al.} 2014,
  \href{http://dx.doi.org/10.1093/mnras/stu1738}{Monthly Notices of the Royal
  Astronomical Society, 445, 581}

\bibitem[{Hunt {et~al.}(2018)Hunt, De~Looze, Boquien, Nikutta, Rossi, Bianchi,
  Dale, Granato, Kennicutt, Silva, Ciesla, Relano, Viaene, Brandl, Calzetti,
  Croxall, Draine, Galametz, Gordon, Groves, Helou, Herrera-Camus, Hinz, Koda,
  Salim, Sandstrom, Smith, Wilson, \& Zibetti}]{Hunt2018}
Hunt, L.~K., De~Looze, I., Boquien, M., {et~al.} 2018,
  \href{http://arxiv.org/abs/1809.04088}{eprint arXiv:1809.04088}

\bibitem[{Hunter(2007)}]{Hunter2007}
Hunter, J.~D. 2007, \href{http://matplotlib.org/citing.html}{Computing in
  Science {\&} Engineering, 9, 90}

\bibitem[{Jones {et~al.}(2001)Jones, Oliphant, Peterson, \&
  {Others}}]{Jones2001}
Jones, E., Oliphant, T., Peterson, P., \& {Others}. 2001, {SciPy: Open source
  scientific tools for Python}

\bibitem[{Kl{\"{o}}ckner {et~al.}(2012)Kl{\"{o}}ckner, Pinto, Lee, Catanzaro,
  Ivanov, \& Fasih}]{Klockner2012}
Kl{\"{o}}ckner, A., Pinto, N., Lee, Y., {et~al.} 2012,
  \href{http://dx.doi.org/10.1016/j.parco.2011.09.001}{Parallel Computing, 38,
  157}

\bibitem[{Kluyver {et~al.}(2016)Kluyver, Ragan-Kelley,
  P{\{}{\textbackslash}'e{\}}rez, Granger, Bussonnier, Frederic, Kelley,
  Hamrick, Grout, Corlay, Ivanov, Avila, Abdalla, \& Willing}]{Kluyver2016}
Kluyver, T., Ragan-Kelley, B., P{\{}{\textbackslash}'e{\}}rez, F., {et~al.}
  2016, in Positioning and Power in Academic Publishing: Players, Agents and
  Agendas, 87

\bibitem[{Krist {et~al.}(2011)Krist, Hook, \& Stoehr}]{Krist2011}
Krist, J.~E., Hook, R.~N., \& Stoehr, F. 2011,
  \href{http://dx.doi.org/10.1117/12.892762}{in Proceedings of the SPIE, Volume
  8127, id. 81270J (2011)., ed. M.~A. Kahan, Vol. 8127}, 81270J

\bibitem[{Kroupa \& {Pavel}(2001)}]{Kroupa2001}
Kroupa, P., \& {Pavel}. 2001,
  \href{http://dx.doi.org/10.1046/j.1365-8711.2001.04022.x}{Monthly Notices of
  the Royal Astronomical Society, Volume 322, Issue 2, pp. 231-246., 322, 231}

\bibitem[{Leja {et~al.}(2018)Leja, Carnall, Johnson, Conroy, \&
  Speagle}]{Leja2018b}
Leja, J., Carnall, A.~C., Johnson, B.~D., Conroy, C., \& Speagle, J.~S. 2018,
  \href{http://arxiv.org/abs/1811.03637}{eprint arXiv:1811.03637}

\bibitem[{Lewis {et~al.}(2015)Lewis, Dolphin, Dalcanton, Weisz, Williams, Bell,
  Seth, Simones, Skillman, Choi, Fouesneau, Guhathakurta, Johnson, Kalirai,
  Leroy, Monachesi, Rix, \& Schruba}]{Lewis2015}
Lewis, A.~R., Dolphin, A.~E., Dalcanton, J.~J., {et~al.} 2015,
  \href{http://dx.doi.org/10.1088/0004-637X/805/2/183}{The Astrophysical
  Journal, 805, 183}

\bibitem[{Lim {et~al.}(2015)Lim, Diaz, \& Laidler}]{Lim2015}
Lim, P.~L., Diaz, R.~I., \& Laidler, V. 2015, {pysynphot: Synthetic photometry
  software package}

\bibitem[{Maraston {et~al.}(2010)Maraston, Pforr, Renzini, Daddi, Dickinson,
  Cimatti, \& Tonini}]{Maraston2010}
Maraston, C., Pforr, J., Renzini, A., {et~al.} 2010,
  \href{http://dx.doi.org/10.1111/j.1365-2966.2010.16973.x}{Monthly Notices of
  the Royal Astronomical Society, 407, 830}

\bibitem[{McKinney(2010)}]{McKinney2010}
McKinney, W. 2010, {Data Structures for Statistical Computing in Python}

\bibitem[{Mould(2012)}]{Mould2012}
Mould, J. 2012, \href{http://dx.doi.org/10.1088/2041-8205/755/1/L14}{The
  Astrophysical Journal, 755, L14}

\bibitem[{P{\'{e}}rez \& Granger(2007)}]{Perez2007}
P{\'{e}}rez, F., \& Granger, B.~E. 2007,
  \href{http://dx.doi.org/10.1109/MCSE.2007.53}{Computing in Science {\&}
  Engineering, 9, 21}

\bibitem[{Pforr {et~al.}(2012)Pforr, Maraston, \& Tonini}]{Pforr2012a}
Pforr, J., Maraston, C., \& Tonini, C. 2012,
  \href{http://dx.doi.org/10.1111/j.1365-2966.2012.20848.x}{Monthly Notices of
  the Royal Astronomical Society, 422, 3285}

\bibitem[{Salpeter(1955)}]{Salpeter1955}
Salpeter, E.~E. 1955, \href{http://dx.doi.org/10.1086/145971}{The Astrophysical
  Journal, 121, 161}

\bibitem[{Schlafly \& Finkbeiner(2011)}]{Schlafly2011}
Schlafly, E.~F., \& Finkbeiner, D.~P. 2011,
  \href{http://dx.doi.org/10.1088/0004-637X/737/2/103}{The Astrophysical
  Journal, 737, 103}

\bibitem[{Skilling(2004)}]{Skilling2004b}
Skilling, J. 2004, \href{http://dx.doi.org/10.1063/1.1835238}{in AIP Conference
  Proceedings, Vol. 735} (AIP), 395

\bibitem[{Sorba \& Sawicki(2015)}]{Sorba2015}
Sorba, R., \& Sawicki, M. 2015,
  \href{http://dx.doi.org/10.1093/mnras/stv1235}{Monthly Notices of the Royal
  Astronomical Society, 452, 235}

\bibitem[{The Astropy~Collaboration {et~al.}(2013)The Astropy~Collaboration,
  Robitaille, Tollerud, Greenfield, Droettboom, Bray, Aldcroft, Davis,
  Ginsburg, Price-Whelan, Kerzendorf, Conley, Crighton, Barbary, Muna,
  Ferguson, Grollier, Parikh, Nair, G{\"{u}}nther, Deil, Woillez, Conseil,
  Kramer, Turner, Singer, Fox, Weaver, Zabalza, Edwards, Bostroem, Burke,
  Casey, Crawford, Dencheva, Ely, Jenness, Labrie, Lim, Pierfederici, Pontzen,
  Ptak, Refsdal, Servillat, \& Streicher}]{TheAstropyCollaboration2013}
The Astropy~Collaboration, A., Robitaille, T.~P., Tollerud, E.~J., {et~al.}
  2013, \href{http://dx.doi.org/10.1051/0004-6361/201322068}{Astronomy {\&}
  Astrophysics, Volume 558, id.A33, 9 pp., 558}

\bibitem[{The Astropy~Collaboration {et~al.}(2018)The Astropy~Collaboration,
  Price-Whelan, Sip{\H{o}}cz, G{\"{u}}nther, Lim, Crawford, Conseil, Shupe,
  Craig, Dencheva, Ginsburg, VanderPlas, Bradley, P{\'{e}}rez-Su{\'{a}}rez,
  de~Val-Borro, Aldcroft, Cruz, Robitaille, Tollerud, Ardelean, Babej,
  Bachetti, Bakanov, Bamford, Barentsen, Barmby, Baumbach, Berry, Biscani,
  Boquien, Bostroem, Bouma, Brammer, Bray, Breytenbach, Buddelmeijer, Burke,
  Calderone, Rodr{\'{i}}guez, Cara, Cardoso, Cheedella, Copin, Crichton,
  D{\'{A}}vella, Deil, Depagne, Dietrich, Donath, Droettboom, Earl, Erben,
  Fabbro, Ferreira, Finethy, Fox, Garrison, Gibbons, Goldstein, Gommers, Greco,
  Greenfield, Groener, Grollier, Hagen, Hirst, Homeier, Horton, Hosseinzadeh,
  Hu, Hunkeler, Ivezi{\'{c}}, Jain, Jenness, Kanarek, Kendrew, Kern,
  Kerzendorf, Khvalko, King, Kirkby, Kulkarni, Kumar, Lee, Lenz, Littlefair,
  Ma, Macleod, Mastropietro, McCully, Montagnac, Morris, Mueller, Mumford,
  Muna, Murphy, Nelson, Nguyen, Ninan, N{\"{o}}the, Ogaz, Oh, Parejko, Parley,
  Pascual, Patil, Patil, Plunkett, Prochaska, Rastogi, Janga, Sabater,
  Sakurikar, Seifert, Sherbert, Sherwood-Taylor, Shih, Sick, Silbiger,
  Singanamalla, Singer, Sladen, Sooley, Sornarajah, Streicher, Teuben, Thomas,
  Tremblay, Turner, Terr{\'{o}}n, van Kerkwijk, de~la Vega, Watkins, Weaver,
  Whitmore, Woillez, Zabalza, Woillez, Zabalza, \&
  Contributors}]{TheAstropyCollaboration2018}
The Astropy~Collaboration, A., Price-Whelan, A.~M., Sip{\H{o}}cz, B.~M.,
  {et~al.} 2018, \href{http://dx.doi.org/10.3847/1538-3881/aabc4f}{The
  Astronomical Journal, Volume 156, Issue 3, article id. 123, 19 pp. (2018).,
  156}

\bibitem[{Tonry \& Schneider(1988)}]{Tonry1988}
Tonry, J., \& Schneider, D.~P. 1988,
  \href{http://dx.doi.org/10.1086/114847}{The Astronomical Journal, 96, 807}

\bibitem[{van~der Walt {et~al.}(2011)van~der Walt, Colbert, \&
  Varoquaux}]{VanderWalt2011}
van~der Walt, S., Colbert, S.~C., \& Varoquaux, G. 2011,
  \href{http://dx.doi.org/10.1109/MCSE.2011.37}{Computing in Science {\&}
  Engineering, 13, 22}

\bibitem[{van Dokkum \& Conroy(2014)}]{vandokkum2014b}
van Dokkum, P., \& Conroy, C. 2014,
  \href{http://dx.doi.org/10.1088/0004-637X/797/1/56}{The Astrophysical
  Journal, Volume 797, Issue 1, article id. 56, 19 pp. (2014)., 797}

\bibitem[{Vogelsberger {et~al.}(2014)Vogelsberger, Genel, Springel, Torrey,
  Sijacki, Xu, Snyder, Nelson, \& Hernquist}]{Vogelsberger2014a}
Vogelsberger, M., Genel, S., Springel, V., {et~al.} 2014,
  \href{http://dx.doi.org/10.1093/mnras/stu1536}{Monthly Notices of the Royal
  Astronomical Society, 444, 1518}

\bibitem[{Walcher {et~al.}(2011)Walcher, Groves, Budav{\'{a}}ri, \&
  Dale}]{Walcher2011}
Walcher, J., Groves, B., Budav{\'{a}}ri, T., \& Dale, D. 2011,
  \href{http://dx.doi.org/10.1007/s10509-010-0458-z}{Astrophysics and Space
  Science, 331, 1}

\bibitem[{Weisz {et~al.}(2014)Weisz, Dolphin, Skillman, Holtzman, Gilbert,
  Dalcanton, \& Williams}]{Weisz2014}
Weisz, D.~R., Dolphin, A.~E., Skillman, E.~D., {et~al.} 2014,
  \href{http://dx.doi.org/10.1088/0004-637X/789/2/147}{The Astrophysical
  Journal, 789, 147}

\bibitem[{Weisz {et~al.}(2011)Weisz, Dalcanton, Williams, Gilbert, Skillman,
  Seth, Dolphin, McQuinn, Gogarten, Holtzman, Rosema, Cole, Karachentsev, \&
  Zaritsky}]{Weisz2011}
Weisz, D.~R., Dalcanton, J.~J., Williams, B.~F., {et~al.} 2011,
  \href{http://dx.doi.org/10.1088/0004-637X/739/1/5}{The Astrophysical Journal,
  739, 5}

\bibitem[{Williams {et~al.}(2018)Williams, Olsen, Khan, Pirone, \&
  Rosema}]{Williams2018a}
Williams, B.~F., Olsen, K., Khan, R., Pirone, D., \& Rosema, K. 2018,
  \href{http://dx.doi.org/10.3847/1538-4365/aab762}{The Astrophysical Journal
  Supplement Series, 236, 4}

\bibitem[{Williams {et~al.}(2015)Williams, Dalcanton, Dolphin, Weisz, Lewis,
  Lang, Bell, Boyer, Fouesneau, Gilbert, Monachesi, \& Skillman}]{Williams2015}
Williams, B.~F., Dalcanton, J.~J., Dolphin, A.~E., {et~al.} 2015,
  \href{http://dx.doi.org/10.1088/0004-637X/806/1/48}{The Astrophysical
  Journal, 806, 48}

\bibitem[{Williams {et~al.}(2017)Williams, Dolphin, Dalcanton, Weisz, Bell,
  Lewis, Rosenfield, Choi, Skillman, \& Monachesi}]{Williams2017}
Williams, B.~F., Dolphin, A.~E., Dalcanton, J.~J., {et~al.} 2017,
  \href{http://arxiv.org/abs/1708.02617}{eprint arXiv:1708.02617}

\end{thebibliography}
